\documentclass[12pt]{article}
\usepackage[margin=1in]{geometry}
\usepackage{graphicx,url}
\usepackage[utf8]{inputenc}  
\usepackage{authblk} % for authors' affiliations

\usepackage{amsmath,amsfonts,amssymb}
\usepackage{natbib}
\usepackage{hyperref}
\usepackage{pdfpages}
\usepackage{array}
\usepackage{dsfont} % to use \mathds{1}

\graphicspath{{figs/}{}}
    
\title{Penalized regression calibration: a method for the prediction of survival outcomes using complex longitudinal and high-dimensional data}

\author[1]{Mirko Signorelli}
\author[2]{Pietro Spitali}
\author[3]{Cristina Al-Khalili Szigyarto}
\author[$ $]{The MARK-MD Consortium}
\author[4]{Roula Tsonaka}

\affil[1]{Mathematical Institute, Leiden University (NL)}
\affil[2]{Department of Human Genetics, Leiden University Medical Center (NL)}
\affil[3]{Department of Protein Science, KTH-Royal Institute of Technology (SW)}
\affil[4]{Department of Biomedical Data Sciences, Leiden University Medical Center (NL)}

\date{}

\begin{document} 

\maketitle

\noindent \textbf{About this article}
\begin{itemize}
\item Please cite this article as: Signorelli, M., Spitali, P., Al-Khalili Szigyarto, C., The MARK-MD Consortium, Tsonaka, R. (2021). Penalized regression calibration: A method for the prediction of survival outcomes using complex longitudinal and high-dimensional data. \textit{Statistics in Medicine}. DOI: 10.1002/sim.9178.
\item This document contains the ``accepted'' version of the manuscript. The final (published) version of the article can be freely downloaded (Open Access) from the website of \textit{Statistics in Medicine}, using this link: \href{https://doi.org/10.1002/sim.9178}{https://doi.org/10.1002/sim.9178}
\end{itemize}

\begin{abstract}
\noindent
Longitudinal and high-dimensional measurements have become increasingly common in biomedical research. However, methods to predict survival outcomes using covariates that are both longitudinal and high-dimensional are currently missing. In this article we propose penalized regression calibration (PRC), a method that can be employed to predict survival in such situations.\\
PRC comprises three modelling steps: first, the trajectories described by the longitudinal predictors are flexibly modelled through the specification of multivariate mixed effects models. Second, subject-specific summaries of the longitudinal trajectories are derived from the fitted mixed models. Third, the time to event outcome is predicted using the subject-specific summaries as covariates in a penalized Cox model.\\
To ensure a proper internal validation of the fitted PRC models, we furthermore develop a cluster bootstrap optimism correction procedure (CBOCP) that allows to correct for the optimistic bias of apparent measures of predictiveness. PRC and the CBOCP are implemented in the \texttt{R} package \texttt{pencal}, available from \texttt{CRAN}.\\
After studying the behaviour of PRC via simulations, we conclude by illustrating an application of PRC to data from an observational study that involved patients affected by Duchenne muscular dystrophy (DMD), where the goal is predict time to loss of ambulation using longitudinal blood biomarkers.

\vspace{0.2cm}
\noindent \textbf{Keywords:} penalized regression calibration; survival analysis; risk prediction modelling; longitudinal data analysis; high-dimensionality; optimism correction; Duchenne muscular dystrophy.
\end{abstract}

\section{Introduction}
\label{sec:intro}

High-dimensionality and longitudinal measurements have become more and more common in biomedical datasets. This evolution has been fostered by several factors, including the increasing availability of clinical registry data and the diffusion of microarray and high-throughput sequencing technologies.
One of the consequences of these changes is that when developing a prediction model for a time to event outcome, biostatisticians are increasingly confronted with the availability of a larger number of possible covariates (often hundreds, or even thousands). More often than not, such covariates can vary over time, leading to scenarios where survival can be predicted using covariates that are both longitudinal and high-dimensional.
Additional features that might need to be dealt with in such scenarios are the presence of potentially strong correlations between the covariates, an unbalanced repeated measurements design, and missing data at certain measurement times for one or more covariates.

An example of such scenarios is represented by the MARK-MD study\citep{signorelli2020,strandberg2020}, an observational study on patients affected by Duchenne muscular dystrophy (DMD) that constitutes the motivating example of this paper. DMD is a rare neuromuscular disorder whose consequences include progressive loss of muscular tissue and muscle mass, loss of ambulation (LoA) around the age of 12, and premature death. To date no cure for DMD has been found, and identification of non-invasive biomarkers that could be employed to monitor disease progression and to predict disease milestones such as LoA is urgent.
The MARK-MD project aimed to study the dynamic evolution of 118 proteins in a longitudinal cohort of DMD patients, and it yielded an unbalanced dataset comprising between 1 and 5 repeated measurements per patient. Protein expression was measured using a suspension bead array platform, whereby the expression levels of a protein are not measured directly, but they are instead implicitly inferred from the relative abundance of multiple antibodies that bind to the protein of interest. In particular, between 1 and 5 antibodies per protein were measured, yielding a total of 240 antibodies matching 118 unique proteins; in most cases, high correlations were found between antibodies that targeted the same protein \citep{signorelli2020}. Statistical analysis of the longitudinal biomarker data \citep{signorelli2020} yielded evidence that several proteins are associated with disease progression and with the wheelchair dependency status, raising the question of whether blood biomarkers may be employed to predict the age at which a clinically meaningful milestone such as LoA will occur.

When developing a model to predict time to LoA for the MARK-MD dataset, three main features should be taken into account: 
\begin{enumerate}
\item \textit{high-dimensionality}: the number of possible predictors (240) largely outweighs the number of subjects (93);
\item the \textit{availability of repeated measurements} (between 1 and 5) from the same subject;
\item the presence of several groups of \textit{strongly correlated predictors}, each of which comprises multiple items (the antibodies) that are measured to reconstruct the same biological process of interest (the protein that the antibodies bind to). 
\end{enumerate}
While several statistical methods are available to tackle separately each of these three features, developing a model that can address them jointly is more challenging.
To date, several approaches to predict survival using a high-dimensional set of cross-sectional covariates have been proposed \citep{vanwieringen2009}, however extensions to high-dimensional settings with longitudinal covariates are currently lacking. On the other hand, study of the association between one or more longitudinal covariates and a time-to-event outcome is typically based on the specification of a joint model for longitudinal and survival data \citep{rizopoulos2012,chen2017}. Estimation of joint models becomes computationally prohibitive when more than a handful of longitudinal covariates are included, restricting their application to low-dimensional settings \citep{hickey2016,mauff2020}. 

In practice, the lack of methods to predict survival using covariates that are both longitudinal and high-dimensional often forces a choice between selecting only a small subset of the available predictors, or using only the baseline measurements of all variables. It is apparent, however, that both choices are suboptimal: if predictions are based on a limited number of pre-specified covariates, important predictors may be missed; if the longitudinal measurements are discarded, important information about the way in which the predictors change over time will be missed. Therefore, a method that avoids these simplifications and does not sacrifice part of the available information would be preferable.

In this article we propose \textit{penalized regression calibration} (PRC), a method that makes it possible to predict survival times using a high-dimensional set of predictors that are measured repeatedly over time, and which may additionally display strong correlations (as in the case of the MARK-MD study). 
PRC comprises three modelling steps, which respectively consist of 1) the specification and estimation of a model for the longitudinal biomarkers, 2) the computation of subject-specific summaries of the trajectories described by the biomarkers, and 3) the estimation of a model for the survival outcome.

More in detail, in step 1 we specify multivariate latent process mixed models (MLPMMs) \citep{proust2013} that allow us to model jointly all the items (i.e., the antibodies) which are employed to measure a latent biological process of interest (the protein targeted by those antibodies). By doing so, we are able to model the trajectories described by highly correlated antibodies using a latent variable model that comprises two sets of random effects, one set for the observed items and another for the latent processes. Moreover, we can easily handle unbalanced designs or missing values thanks to the flexibility of mixed effects models. In step 2 we proceed to the computation of subject-specific summaries of the longitudinal biomarkers, which in practice are the predicted random effects from the MLPMM. In step 3 we employ these summaries as predictors of the time to event outcome (time to LoA in the MARK-MD study) by estimating a penalized Cox model with ridge or elasticnet penalty, which can then be used to compute predictions of survival probabilities. 

In addition to this first model based on the MLPMM, we also consider a simpler model where in step 1 the biomarkers are modelled independently using linear mixed models (LMMs) \citep{mcculloch2008}. Such an approach, which performs univariate modelling of the biomarkers and is thus computationally less intensive, might be preferred in situations where no distinction between observed items and latent biological processes of interest exists. In terms of relationship with existing statistical methods, PRC might be regarded as a high-dimensional extension of the ordinary regression calibration \citep{ye2008} (ORC) approach. However, ORC and PRC differ in their purposes: while the goal of ORC is to estimate the \textit{association} between a single longitudinal biomarker and survival, PRC aims to \textit{predict} survival based on a high-dimensional set of longitudinal biomarkers.

\begin{figure}
\centering
\includegraphics[scale=0.4]{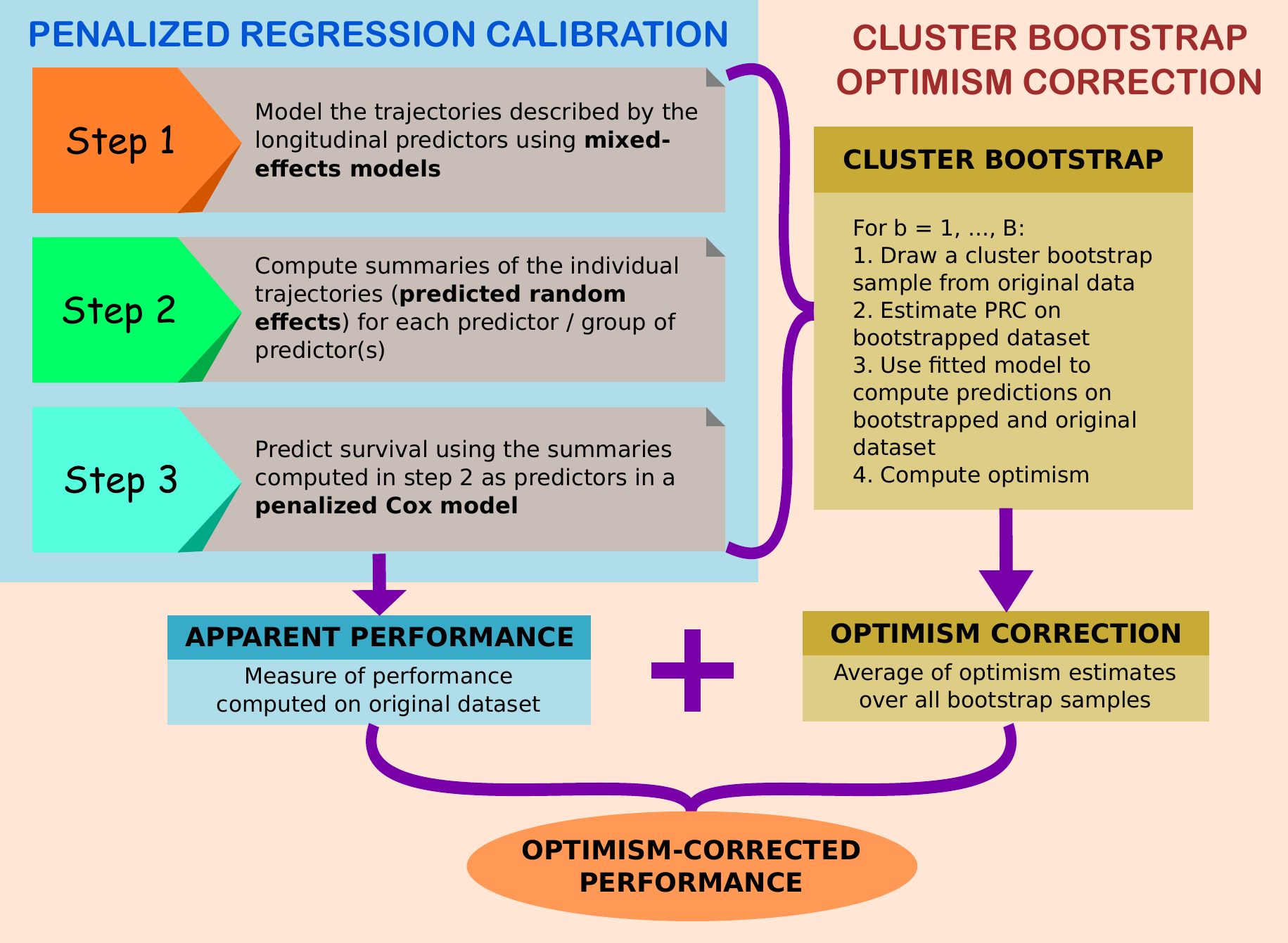}
\caption{Schematic representation of PRC and CBOCP. The lightblue box summarizes the three modelling steps that PRC consists of. The faded orange box shows the steps involved in the computation of the CBOCP.}
\label{fig:prc_diagram}
\end{figure}

A further contribution of our work is the implementation of a computational procedure that allows to perform the internal validation of PRC. The internal validation of prediction models based on penalized regression is often a neglected step, possibly due to the belief that the penalization of the model parameters will suffice to prevent overfitting, so that no formal assessment of the optimism associated to the reported performance is then needed. In this paper we do not take this belief for granted, and instead we explicitly explore the topic of the internal validation of PRC, which is a penalized prediction model. In particular, we implement a \textit{cluster bootstrap optimism correction procedure} (CBOCP hereafter) that allows to estimate the optimism of naive predictiveness measures, thereby making it possible to properly assess whether overfitting is present. The CBOCP requires the re-estimation of the PRC model on a sequence of clustered bootstrap samples, and the computation of estimates of the optimism associated to the apparent model performance, which are obtained by comparing the predictive performance of the newly fitted models on the bootstrapped dataset and on the original dataset. When averaged over all cluster bootstrap samples, these estimates yield a bootstrap optimism correction that is subtracted from the apparent measure of model predictiveness (i.e., the measure of predictiveness computed on the original dataset where PRC was fitted). 
In Figure \ref{fig:prc_diagram} we provide a schematic representation of the different steps involved in the estimation of PRC, and in the computation of the CBOCP.

Lastly, we provide an implementation of functions to compute PRC and its associated CBOCP using \verb|R|, and make them available through the \verb|R| package \verb|pencal| \citep{signo-pencal}, which can be freely downloaded from \verb|CRAN|.

The remainder of the article is organized as follows: in Section \ref{sec:methods} we introduce the methodological framework for PRC, discussing in detail the three steps involved in its estimation and the CBOCP. In Section \ref{sec:sims} we employ simulations to compare PRC to a simpler prediction approach that employs information on the baseline biomarker levels, but ignores the repeated measurements taken after baseline; 
additionally, we compare the predictive performance of PRC to that of a joint modelling approach in a low-dimensional setting with 3 longitudinal covariates.
In Section \ref{sec:appl} we employ PRC to predict time to LoA for the patients involved in the MARK-MD study. Lastly, in Section \ref{sec:concl} we provide some conclusions and discuss possible extensions of PRC.

\section{Methods}
\label{sec:methods}

We consider a situation in which a set of $p$ latent biological processes (indexed by $s \in \{1, ..., p \}$) are considered as potential predictors of an event of interest, but they are not measured directly. Instead, $r_s \geq 1$ items (indexed by $q \in \{1, ..., r_s\}$) that are related to the $s$-th biological process are employed to reconstruct it, so that in total, $\sum_{s=1}^p r_s$ items are employed as proxies for the $p$ latent biological processes. In simpler situations where each biological process of interest can be measured directly, or where $r_s = 1$ for all $s \in \{1, ..., p\}$, the distinction between items and latent processes is unnecessary, and the subscript $q$ can be dropped.

The distinction between a latent biological process of interest and the items used to measured it is motivated by the setup of proteomic studies, where typically the expression of a protein is not measured directly, but it is instead reconstructed by measuring the abundance of multiple antibodies that bind to the protein itself.
In the case of the MARK-MD study, the latent biological processes are $p = 118$ proteins, and the items are the $\sum_{s=1}^p r_s = 240$ antibodies that were measured to reconstruct the level of the proteins. In particular, between 1 and 5 antibodies were measured for each protein, so that $r_s \in \{1, 2, 3, 4, 5\}$.
This distinction is not unique to proteomic studies: it may as well arise in different settings, for example when different psychometric items are employed to reconstruct a latent cognitive process of interest.

We assume that the items are measured longitudinally following either a balanced or an unbalanced design, and that for each individual $i \in \{1, ..., n\}$, $m_i \geq 1$ repeated measurements are collected before the event of interest occurrs. We denote by $y_{qsij}$ the value of the $q$-th item related to the $s$-th latent biological process that is measured on individual $i$ at their $j$-th repeated measurement ($j \in \{1, ..., m_i\}$). We summarize all $\sum_{i=1}^n m_i$ measurements of the $q$-th item referring to the $s$-th biological process in a vector $y_{qs} = (y_{qs11}, ..., y_{qsn m_n})$. 

Moreover, we use $a_{ij}$ to denote the age of subject $i$ when the $j$-th measurement is collected, and we let $t_i > 0$ be the event time (measured starting from the baseline age $a_{i1}$) for individual $i$, and $\delta_i$ be a censoring indicator such that $\delta_i = 1$ if the event of interest is observed at $t_i$, or $\delta_i = 0$ in case of right-censoring. 

Let $y_i = (y_{11i1}, ..., y_{r_s s i m_i})$ be the vector containing all $m_i$ repeated measurements of all items obtained from subject $i$. In Sections \ref{sub:prc-step1}, \ref{sub:prc-step2} and \ref{sub:prc-step3} we propose PRC, a method that allows to predict the time to event outcome $t_i$ using as predictor $y_i$, which comprises a high-dimensional set of longitudinally measured items.
Specifically, in Section \ref{sub:prc-step1} we model the trajectories described by the longitudinal biomarkers using the MLPMM \citep{proust2013}. In Section \ref{sub:prc-step2} we proceed to the computation of subject-specific summaries of the longitudinal trajectories that refer both to the measured items, and to the latent biological process. In Section \ref{sub:prc-step3} we specify a penalized Cox model that relates the time to event outcome to these summaries, and we employ it to compute predictions of survival probabilities.

In Section \ref{sub:bootstrap} we turn our attention to the problem of correctly assessing the predictive performance of PRC, discussing the CBOCP that we have implemented to obtained unbiased estimates of in-sample predictiveness. Lastly, in Section \ref{sub:code} we provide details about the implementation of the proposed methodology in \verb|R|, and we shortly introduce the \verb|R| package \verb|pencal| \citep{signo-pencal}.

\subsection{Modelling the longitudinal biomarkers}
\label{sub:prc-step1}

The first step of PRC involves the estimation and specification of a statistical model capable of describing the evolution over time of the longitudinal biomarkers. 
We propose to jointly model the dynamic evolution of all the items $\left(y_{1s}, y_{2s}, ..., y_{r_s s}\right)$ that refer to the $s$-th latent biological process with the following MLPMM \citep{proust2013}:
\begin{equation}
y_{qsij} = \beta_{qs0} + u_{s0i} + b_{qsi} + \beta_{qs1} a_{ij} + u_{s1i} a_{ij} + \varepsilon_{qsij}, \:\: q = 1, ..., r_s,
\label{eq:mlcmm}
\end{equation}
where $\beta_{qs0}$ and $\beta_{qs1}$ are fixed effect parameters, $u_{si} = (u_{s0i}, u_{s1i}) \sim N_2 (0, \Sigma_{u_s})$ is a vector of \textit{shared random effects} that comprises a shared random intercept ($u_{s0i}$) and a shared random slope ($u_{s1i}$), $b_{qsi} \sim N_1(0, \sigma^2_{b_{qs}}) \; \forall q \in \{1, ..., r_s\}$ are \textit{item-specific random intercepts}, and $\varepsilon_{qsij} \sim N_1(0, \sigma^2_{\varepsilon_{qs}})$ is a measurement-error term. As identifiability condition, we set $Var(u_{s1i}) =1$. If there are some processes that are measured using only one item, i.e. there exist some $s$ for which $r_s = 1$, the MLPMM of equation \eqref{eq:mlcmm} can be simplified into $y_{sij} = \beta_{s0} + u_{s0i} + \beta_{s1} a_{ij} + u_{s1i} a_{ij} + \varepsilon_{sij}$, which is a LMM with correlated random intercept and random slope.

We employ maximum likelihood to estimate model \eqref{eq:mlcmm}, using the \verb|R| package \verb|lcmm| \citep{proust2017}. Note that for simplicity we have included only age as covariate and random slope in model \eqref{eq:mlcmm}, but depending on the features of the problem at hand one may choose to add to model \eqref{eq:mlcmm} further covariates (for example to account for relevant confounders), and to include further shared random effects.

The advantage of using the MLPMM specified in \eqref{eq:mlcmm} is two-fold. 
First, it allows us to model jointly all the items $\left(y_{1s}, ..., y_{r_ss} \right)$ that refer to the $s$-th latent biological process, so that we can properly account for the possibility that the items may be strongly correlated with each other. 
Second, the MLPMM enables us to separate the within-item variability from the between-item variability, thanks to the inclusion of a first set of random effects that refer to the observed items, and of a second set of random effects that are linked to the unobserved biological process of interest. In practice, this means that although we cannot directly measure the relevant biological process, we can describe its underlying dynamic evolution using the shared random effects in \eqref{eq:mlcmm}.

This multivariate modelling approach is motivated by the setup of the MARK-MD study, where proteins were not measured directly, but through multiple antibodies, and it can be applied to any other situation in which one measures multiple items that can be related to a common latent process. While our framework is general enough to account for such complex designs, it can be flexibly adapted to simpler situations where the biological processes of interest are directly measured. In such scenarios, one might prefer to model the longitudinal biomarkers independently from each other using separate LMMs \cite{mcculloch2008}. 
Since in such cases $r_s = 1$ for all $s = 1,..., p$, we can drop the subscript $q$ from $y_{qsij}$, and simplify the notation to $y_{sij}$. The general formulation of the LMM in matrix notation is
\begin{equation}
y_{si} = X_i \beta_s + Z_i b_{si} + \varepsilon_{si},
\label{eq:lmm-matr}
\end{equation}
where $X_i$ and $Z_i$ are design matrices that respectively refer to the fixed effects coefficients $\beta_s$ and to the random effects $b_{si}$, $b_{si} \sim N(0, D_s)$ is a vector of normally-distributed random effects, and $\varepsilon_{si} \sim N(0, \sigma^2_s I_{m_i} )$ is a gaussian error term.

The LMM framework makes it possible to specify a wide range of random effect structures in \eqref{eq:lmm-matr}; in particular, to keep the analogy with model \eqref{eq:mlcmm} we hereafter consider the following LMM with correlated random intercepts and random slopes:
\begin{equation}
y_{sij} = \beta_{s0} + b_{s0i} + (\beta_{s1} + b_{s1i}) a_{ij} + \varepsilon_{sij},
\label{eq:lmm}
\end{equation}
where $b_{si} = (b_{s0i}, b_{s1i}) \sim N_2(0, D_{s})$ comprises a random intercept $b_{s0i}$ and a random slope $b_{s1i}$, and $\varepsilon_{si} \sim N_{m_i}(0, \sigma^2_{\varepsilon_s} I_{m_i})$. 
We employ maximum likelihood for the estimation of model \eqref{eq:lmm}, resorting in particular to the \verb|R| package \verb|nlme| \citep{pinheiro2020}.

\subsection{Derivation of the predicted random effects}
\label{sub:prc-step2}

In the second step of PRC we compute summaries of the longitudinal trajectories described by the biomarkers based on the mixed models estimated in the previous step. These summaries are the predicted random effects \citep{tsonaka2012}, and their computation (and interpretation) differs depending on the type of model fitted in the first step. 

Computation of the predicted random effects in \verb|R| is typically straightforward when performed for the very same dataset on which the mixed effects model was fitted. However, it is more complicated when one needs to compute the predicted random effects for a new dataset, because this is not implemented neither in the \verb|lcmm| package that estimates the MLPMM, nor in the popular \verb|R| packages \verb|nlme| and \verb|lme4| that are commonly used to estimate LMMs. 
Nevertheless, the ability to compute the predicted random effects on a dataset different from that on which a MLPMM or a LMM is fitted is fundamental for the implementation of the optimism correction that we will describe in Section \ref{sub:bootstrap}, as well as for the computation of predictions for a new individual that was not included in the original dataset on which PRC was fitted. Therefore, hereafter we illustrate how to compute the predicted random effects for the MLPMM and for the LMM.

For the MLPMM, we can compute the predicted random effects $\hat{u}_{si} = (\hat{u}_{s0i}, \hat{u}_{s1i})$ and $\hat{b}_{si} = (\hat{b}_{1si}, ..., \hat{b}_{r_ssi})$ by adapting the formulas provided by Ebrahimpoor and coauthors \citep{ebrahimpoor2021} to the MLPMM of equation \eqref{eq:mlcmm} as follows:

\begin{equation}
\left( \hat{u}_{si}, \hat{b}_{si} \right) =
E \left( u_{si}, b_{si} |Y_{si} = y_{si} \right) = \left[ 
\begin{matrix}
Z_i \hat{\Sigma}_{u_s}\\
\hat{\Sigma}_{b_s} I_{r_s} \otimes \mathds{1}_{m_i, 1}
\end{matrix}
 \right] 
 \hat{\Sigma}_{y_{si}}^{-1} \dot{y}_{si},
\label{eq:mlpmm-ranef}
\end{equation}

\noindent where $y_{si} = (y_{1si1}, ..., y_{1sim_i}, ..., y_{r_ssi1}, ..., y_{r_ssim_i})^T$, $\dot{y}_{si}$ is the equivalent of $y_{si}$ with $\dot{y}_{qsij} = y_{qsij} - \hat{\beta}_{qs0} - \hat{\beta}_{qs1} a_{ij}$ as entries, $Z_i$ is the random-effects design matrix associated to $y_{si}$ in \eqref{eq:mlcmm},
$\Sigma_{b_s} = \left[ 
\begin{matrix}
\sigma^2_{b1s} & ... & 0\\ 
\vdots & \ddots & \vdots\\
0 & ... & \sigma^2_{br_ss}\\ 
\end{matrix}
\right]$,
$\Sigma_{\varepsilon_s} = \left[ 
\begin{matrix}
\sigma^2_{\varepsilon 1s} & ... & 0\\ 
\vdots & \ddots & \vdots\\
0 & ... & \sigma^2_{\varepsilon r_s s}\\ 
\end{matrix}
\right]$,
$\Sigma_{u_s} = \left[ 
\begin{matrix}
\sigma^2_{us0} & \sigma_{us0, us1}\\ 
\sigma_{us0, us1} & \sigma^2_{us1}
\end{matrix}
\right]$ and
$ \Sigma_{y_{si}} = Z_i \Sigma_{u_s} Z_i^T + I_{r_s} \otimes \Sigma_{\varepsilon_s} I_{m_i}
+ I_{r_s} \otimes \Sigma_{b_s} \mathds{1}_{m_i, m_i}, $
where $I$ denotes identity matrices and $\mathds{1}_{a,b}$ ``all-ones'' matrices (i.e., matrices whose entries are all equal to 1) of dimension $a \times b$.

For the LMM, instead, the predicted random effects are given by
\begin{equation}
\hat{b}_{si} = E(b_{si} | Y_{si} = y_{si}) = \hat{D}_{s} Z_i^T \hat{V}_{si}^{-1} (y_{si} - X_i \hat{\beta}_{s}), 
\label{eq:lmm-ranef}
\end{equation}
where $V_{si} = Z_i D_{s} Z_i^T + \sigma^2_{\varepsilon_s} I_{m_i}$ is the marginal covariance matrix of subject $i$. 

\subsection{Prediction of the survival outcome}
\label{sub:prc-step3}

In the last step of the PRC approach we employ the summary measures obtained in Section \ref{sub:prc-step2} to compute predictions of the individual survival probabilities. To achieve this goal, we specify a penalized Cox model where we include baseline age and the predicted random effects as covariates. 
When the longitudinal biomarkers are modelled using the MLPMM of equation \eqref{eq:mlcmm}, we can consider two alternative prediction approaches.

Let $\hat{u}_{i} = \left(\hat{u}_{10i}, ..., \hat{u}_{p1i} \right)$ and $\hat{b}_{i} = (\hat{b}_{11i}, ..., \hat{b}_{r_ppi})$ denote the vectors of shared and item-specific predicted random effects for subject $i$ computed using \eqref{eq:mlpmm-ranef}. The first approach that we consider is proposed having in mind a data generating mechanism where the event times are influenced only by the latent biological processes: with such an approach, the items are viewed just as an instrument to reconstruct the unobservable biological process, and they are not believed to have an additional predictive value of their own. Thus, in this first approach the hazard function depends only on the shared random effects:
\begin{equation}
h(t_i | a_{i1}, \hat{u}_{i}) = h_0(t_i) \exp \left( \tau a_{i1} 
+ \sum_{s=1}^p \gamma_s \hat{u}_{s0i} + \sum_{s=1}^p \delta_s \hat{u}_{s1i} \right),
\label{eq:prc-mlpmmu}
\end{equation}
where $h(t_i | a_{i1}, \hat{u}_{i})$ denotes the hazard function, and  $h_0(t_i)$ is the baseline hazard. Hereafter, we refer to this modelling approach as \textit{PRC MLPMM(U)}.

On the contrary, the second approach that we consider allows for the possibility that besides being useful to reconstruct the latent biological processes, the items may carry additional information that is useful for the prediction of the time to event outcome. Thus, in this second approach the item-specific random effects $\hat{b}_{i}$ are included as predictors alongside with the shared $\hat{u}_{i}$:
\begin{equation}
h(t_i | a_{i1}, \hat{u}_{i}, \hat{b}_{i}) = h_0(t_i) \exp \left( \tau a_{i1} 
+ \sum_{s=1}^p \gamma_s \hat{u}_{s0i} + \sum_{s=1}^p \delta_s \hat{u}_{s1i} 
+ \sum_{s=1}^p \sum_{q=1}^{r_s} \xi_{qs} \hat{b}_{qsi} \right).
\label{eq:prc-mlpmmu+b}
\end{equation}
We refer to this modelling approach as \textit{PRC MLPMM(U+B)}.

Lastly, in the case in which the longitudinal markers are modelled with univariate LMMs we let
\begin{equation}
h(t_i | a_{i1}, \hat{b}_{i}) = h_0(t_i) \exp \left( \tau a_{i1} 
+ \sum_{s=1}^p \gamma_{s} \hat{b}_{s0i} + \sum_{s=1}^p \delta_{s} \hat{b}_{s1i} \right),
\label{eq:prc-lmm}
\end{equation}
where $\hat{b}_{i} = (\hat{b}_{101}, \hat{b}_{111}, ..., \hat{b}_{p0i}, \hat{b}_{p1i})$ is the vector containing all the predicted random effects for subject $i$, as computed with \eqref{eq:lmm-ranef}. We refer to this modelling approach as \textit{PRC LMM}.

In models \eqref{eq:prc-mlpmmu}, \eqref{eq:prc-mlpmmu+b} and \eqref{eq:prc-lmm} we have included baseline age as covariate, assuming that subjects entered the study at different ages. If this is not the case, $a_{i1}$ can be dropped. Furthermore, in practice one may want or need to include in such models relevant predictors or confounders that are not measured longitudinally, such as for example gender, ethnicity, hospital, etc.; obviously, this can be easily done by adding such covariates to the linear predictors.

Models \eqref{eq:prc-lmm}, \eqref{eq:prc-mlpmmu} and \eqref{eq:prc-mlpmmu+b} may (and typically will) comprise a large, high-dimensional set of predictors. For example, in the MARK-MD study the number of predictors is 237 for the PRC MLPMM(U) approach of equation \eqref{eq:prc-mlpmmu}, 355 for the PRC MLPMM(U+B) in \eqref{eq:prc-mlpmmu+b}, and 481 for the PRC LMM approach in \eqref{eq:prc-lmm}. 
In such high-dimensional scenarios, maximum likelihood estimation is unfeasible, and some form of regularization is needed. 
In this article we consider two penalized likelihood estimation methods, respectively based on the ridge \citep{verweij1994} and on the elasticnet \citep{simon2011} penalties. 
The elasticnet penalty is a linear combination of an $L^1$ and an $L^2$ penalty that comprises two tuning parameters $\alpha \in [0, 1]$ and $\lambda > 0$. $\lambda$ determines the overall level of regularization imposed on the regression coefficients by the penalty, whereas $\alpha$ determines the relative contributions of the two penalties. When $\alpha = 0$, the ridge penalty is obtained, whereas when $\alpha = 1$, we obtain the lasso penalty.
For model \eqref{eq:prc-mlpmmu+b}, the elasticnet penalty is given by
\begin{equation}
p(\gamma, \delta, \xi; \lambda, \alpha) = \lambda \left[
\alpha \left( \sum_{s=1}^p |\gamma_s| + \sum_{s=1}^p |\delta_s| + \sum_{s=1}^p \sum_{q=1}^{r_s} |\xi_{qs}| \right) +
(1-\alpha) \left( \sum_{s=1}^p \gamma_s^2 + \sum_{s=1}^p \delta_s^2 + \sum_{s=1}^p \sum_{q=1}^{r_s} \xi_{qs}^2 \right)
 \right],
\label{eq:elnet}
\end{equation}
where we do not penalize the regression coefficient associated to the baseline age. The ridge penalty can be obtained from \eqref{eq:elnet} by fixing $\alpha = 0$. For models \eqref{eq:prc-mlpmmu} and \eqref{eq:prc-lmm}, the penalties can be obtained from equation \eqref{eq:elnet} by dropping the terms containing $\xi_{qs}$.   

We employ the \verb|R| package \verb|glmnet| \citep{simon2011} to estimate the penalized Cox models described above. We use the cross-validation procedure implemented in \verb|glmnet| to select the optimal value of tuning parameter $\lambda$ when the ridge penalty is used. Moreover, we implement a nested cross-validation procedure \citep{engebretsen2019} to select the optimal values of the two tuning parameters $(\lambda, \alpha)$ when the elasticnet penalty is used. 

Once the penalized Cox model of step 3 is estimated, we can proceed with the computation of the predicted survival probabilities. As usual with the Cox model, these can be obtained as follows:
\begin{equation*}
\hat{S_i}(t) = \exp \left( - \int_0^t \hat{h}_0(z) e^{\hat{\eta}_{ij}} dz \right),
\end{equation*}
where $\hat{h}_0(z)$ is a non-parametric estimate of the baseline hazard function, and $\eta_{ij}$ denotes the linear predictor in models \eqref{eq:prc-mlpmmu}, \eqref{eq:prc-mlpmmu+b}, and \eqref{eq:prc-lmm}. 
Conditional survival probabilities can be obtained as $\hat{S_i}(t_2 | t_1) = \frac{\hat{S_i}(t_2)}{\hat{S_i}(t_1)}$, where $t_2 > t_1$.

\subsection{Cluster bootstrap optimism-correction procedure (CBOCP)}
\label{sub:bootstrap}

We employ two measures of predictiveness to evaluate the predictive performance of the PRC-LMM and PRC-MLPMM methods: the time-dependent area under the ROC curve (tdAUC hereafter), which we estimate using the nearest neighbour estimator proposed by Heagerty and coauthors \citep{heagerty2000}, and the concordance (or C) index, which we estimate using the method proposed by Pencina and coauthors\citep{pencina2004}.

A common problem when developing prediction models is that of correctly evaluating the predictive performance of a method. Typically, the naive estimate of an index of predictiveness will be optimistically biased, due to the overfitting caused by the fact that the naive measure of performance is computed using the same data on which the model is estimated.
For this reason, with low-dimensional problems it is customary to implement an internal validation strategy that allows to correct for this bias\citep{steyerberg2009}. However, internal validation is often a neglected step in the development of high-dimensional risk prediction models, probably due to the belief that the penalization introduced in the estimation phase will by itself prevent any overfitting. In this paper we do not subscribe to this belief, and instead we develop a CBOCP that makes it possible to estimate the bias of the naive measures of performance, thus quantify the extent of the overfitting.

Our CBOCP adapts the boostrap optimism correction method \citep{efron1983} to the case of longitudinal data. Hereafter we illustrate how the correction is computed in the case of the C index, but the computation proceeds in the same manner for the tdAUC. Let $s_0 = \{ 1,..., n\}$ be the set of subject indicators contained in the dataset at hand, which we denote by $D(s_0)$. Note that for each subject $i$, $D(s_0)$ contains $m_i$ repeated measurements. The CBOCP proceeds through the following steps:

\begin{enumerate}
\item draw from $s_0$ a sample with replacement comprising $n$ (possibly replicated) subject ids. Denote by $s_b$ the bootstrap sample thus obtained, and by $D(s_b)$ the corresponding longitudinal dataset;
\item estimate the PRC model of interest using $D(s_b)$ as data. Denote by $\hat{M}_b$ the model thus fitted;
\item estimate the value of the C index of $\hat{M}_b$ when applied to $D(s_b)$, $\hat{C}_b$, and to the original dataset $D(s_0)$, $\hat{C}_{0b}$;
\item repeat steps (1-3) $b = 1, ..., B$ times (typically, $B = 100$ or $200$) 
\item compute the optimism correction $\frac{1}{B} \sum_{b=1}^B (\hat{C}_b - \hat{C}_{0b})$.
\end{enumerate}

Let $\hat{C}_0$ denote the naive estimate of the C index. The optimism-corrected C index, $\hat{C}_{CBOCP}$, is obtained by substracting the optimism correction from the naive C index:
\begin{equation}
\hat{C}_{CBOCP} = \hat{C}_0 - \frac{1}{B} \sum_{b=1}^B (\hat{C}_b - \hat{C}_{0b}).
\end{equation}

\subsection{The R package pencal}
\label{sub:code}

We have implemented the methodology described in Sections \ref{sub:prc-step1}, \ref{sub:prc-step2}, \ref{sub:prc-step3} and \ref{sub:bootstrap} in the \texttt{R} package \texttt{pencal} \citep{signo-pencal}, which can be freely downloaded from \texttt{CRAN} at the link \url{https://cran.r-project.org/package=pencal}. The package comprises functions to estimate the PRC MLPMM(U), PRC MLPMM(U+B) and PRC LMM models, to compute the CBOCP, and to obtain the predicted survival probabilities from the fitted models. A vignette illustrating how to use the package is also available on \texttt{CRAN}. The scripts used for the simulations described in Section \ref{sec:sims} are available at \url{https://github.com/mirkosignorelli/pencal}.

\section{Simulations}
\label{sec:sims}

As mentioned in Section \ref{sec:intro}, the current lack of methods that can predict survival using predictors that are both longitudinal and high-dimensional forces the use of simpler prediction strategies, which either ignore the longitudinal information, or focus on a limited number of predictors and ignore the remaining ones. 
In this section we describe the results of simulations studies designed to compare PRC to a simpler prediction approach where the baseline biomarker levels are employed as predictors in a penalized Cox model. This approach, which we refer to as \textit{baseline pCox} model hereafter, can handle the large number of available biomarkers, but it ignores the repeated measurements that carry information on their evolution over time. In Sections \ref{sub:lmmsims} and \ref{sub:mlpmmsims} we compare the baseline pCox approach to the PRC LMM, PRC MLPMM(U) and PRC MLPMM(U+B) methods, showing that by properly modelling the dynamic evolution of the biomarkers we can improve the accuracy of predictions in a high-dimensional setting. In Section \ref{sub:naivevscboc} we compare the distribution of the naive and optimism-corrected C index and tdAUC, showing the importance of the CBOCP in evaluating overfitting (and correcting for it). 
Lastly, in Section \ref{sub:comptime} we turn our attention to computing time.

\subsection{Evaluation of the PRC LMM and the effect of the number of repeated measurements}
\label{sub:lmmsims}

\begin{figure}
\centering
\includegraphics[scale=0.35, page = 1]{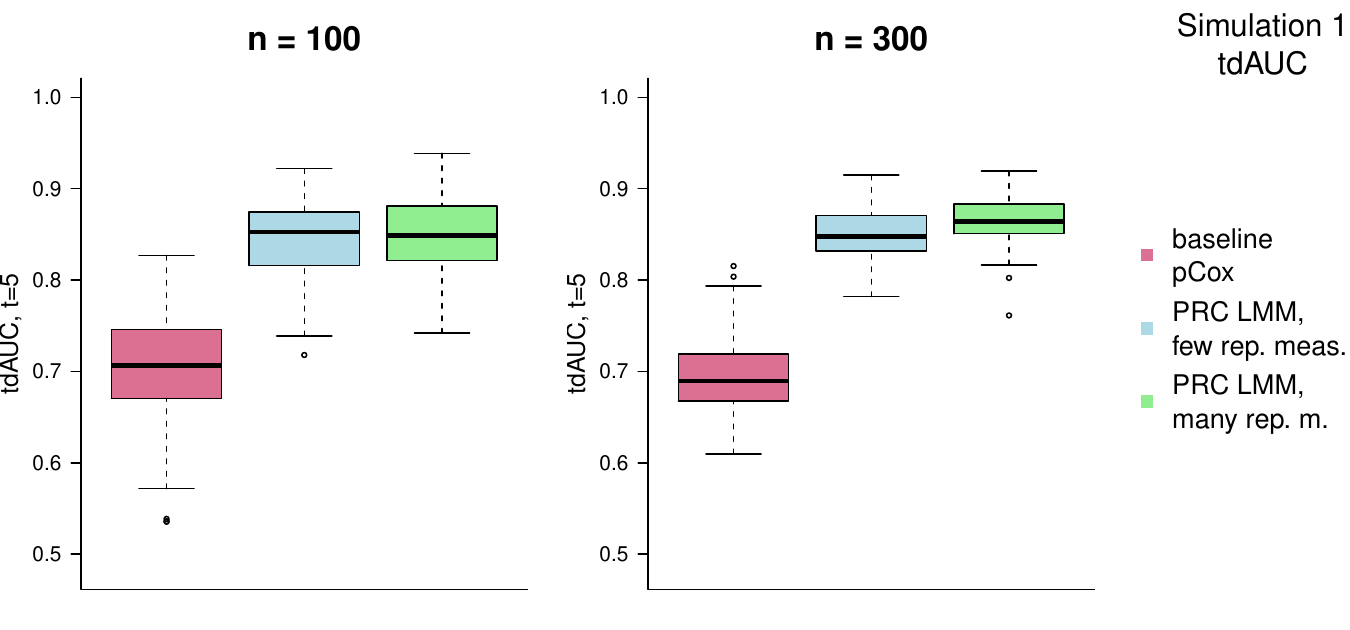}
\includegraphics[scale=0.35, page = 1]{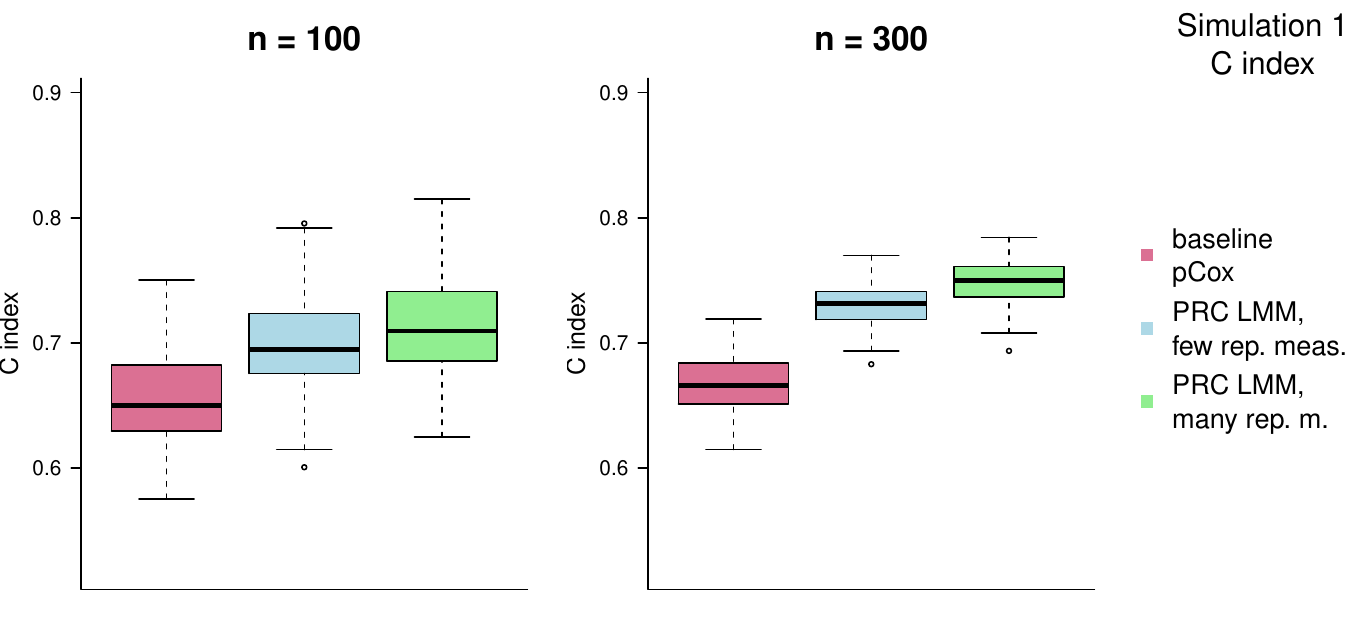}\\
\includegraphics[scale=0.35, page = 2]{tdAUC_univ_ridge_lowd.pdf}
\includegraphics[scale=0.35, page = 2]{Cindex_univ_ridge_lowd.pdf}\\
\includegraphics[scale=0.35, page = 3]{tdAUC_univ_ridge_lowd.pdf}
\includegraphics[scale=0.35, page = 3]{Cindex_univ_ridge_lowd.pdf}
\caption{Results of simulations 1, 2 and 3 (where we consider a low-dimensional setting with 30 biomarkers) using the ridge penalty. The boxplots compare the distribution over 100 random replications of the optimism-corrected tdAUC (left) and C index (right) of the PRC LMM model when few (lightblue) or many (lightgreen) repeated measurements are available to that of a penalized Cox model where only baseline measurements are used (violet red).}
\label{fig:sim1}
\end{figure}

\begin{figure}
\centering
\includegraphics[scale=0.35, page = 1]{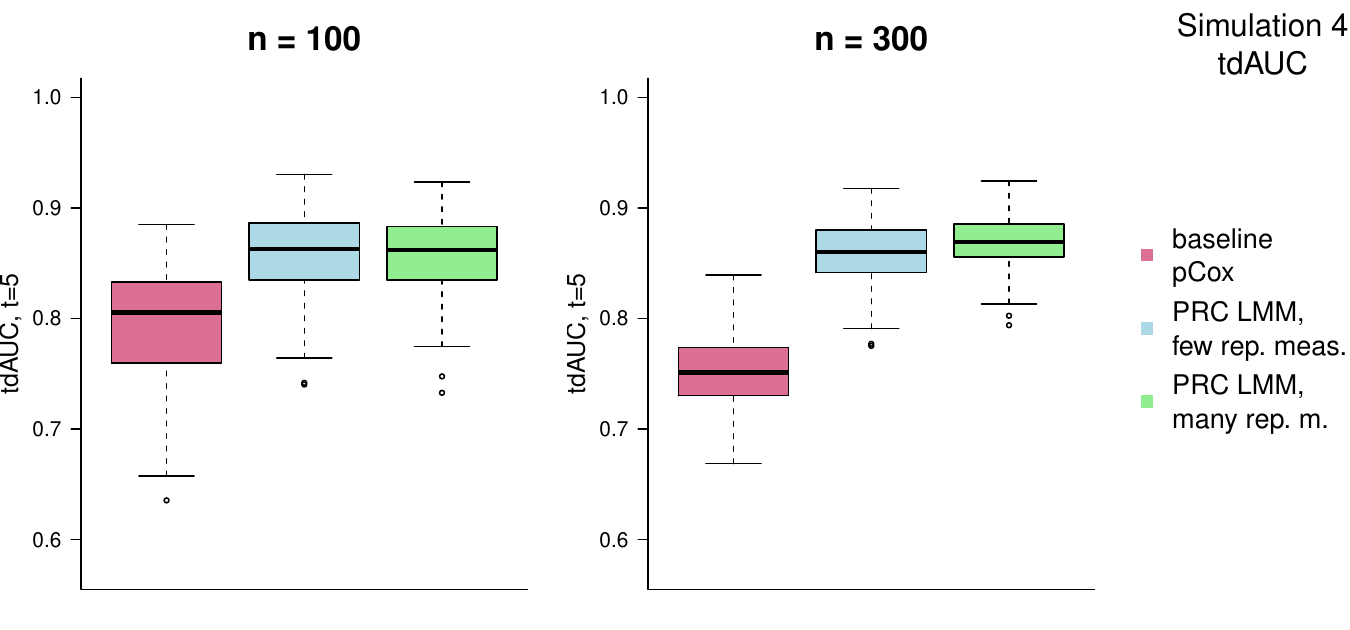}
\includegraphics[scale=0.35, page = 1]{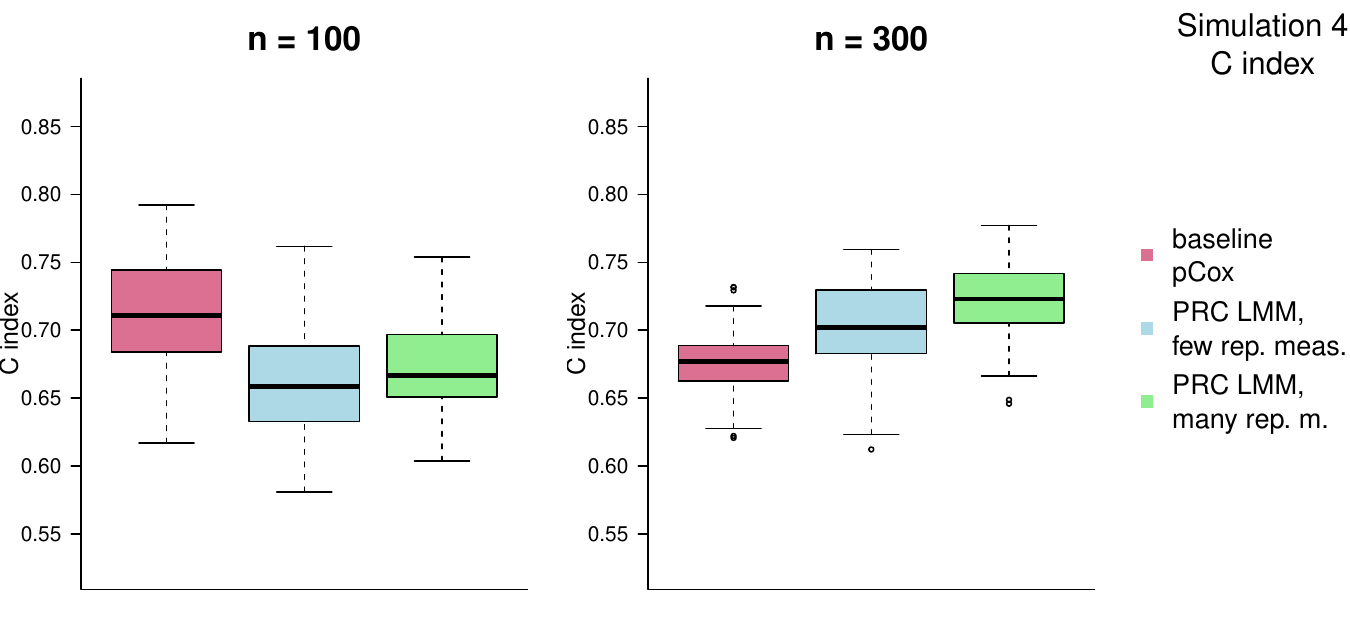}\\
\includegraphics[scale=0.35, page = 2]{tdAUC_univ_ridge_highd.pdf}
\includegraphics[scale=0.35, page = 2]{Cindex_univ_ridge_highd.pdf}\\
\includegraphics[scale=0.35, page = 3]{tdAUC_univ_ridge_highd.pdf}
\includegraphics[scale=0.35, page = 3]{Cindex_univ_ridge_highd.pdf}
\caption{Results of simulations 4, 5 and 6 (where we consider a high-dimensional setting with 150 biomarkers) using the ridge penalty. The boxplots compare the distribution over 100 random replications of the optimism-corrected tdAUC (left) and C index (right) of the PRC LMM model when few (lightblue) or many (lightgreen) repeated measurements are available to that of a penalized Cox model where only baseline measurements are used (violet red).}
\label{fig:sim4}
\end{figure}

We begin our simulation studies by evaluating how the level of improvement in predictive accuracy that can be obtained using PRC can depend on the number of available repeated measurements and on the type of relationship existing between the longitudinal biomarker profiles and survival. For simplicity, here we consider a situation where no distinction between items and latent biological processes is present (namely, $r_s = 1 \; \forall s = 1, ..., p$), and we compare the PRC LMM and baseline pCox approaches. In Section \ref{sub:mlpmmsims} we will consider more complex scenarios where such distinction is present, and we will employ the PRC MLPMM instead of the PRC LMM.

We first consider a low-dimensional setting with $p = 30$ biomarkers. Without loss of generality, we assume that all patients entered the study at the same age. We generate the longitudinal profiles of each biomarker from the LMM of equation \eqref{eq:lmm}, considering two cases that differ for the number of available repeated measurements per patient (we will refer to these two cases as scenarios with ``few'' and ``many'' repeated measurements), and we simulate survival times from a Weibull model \citep{bender2005}. We consider 3 simulations: in simulation 1 we assume that the random intercepts and random slopes are uncorrelated and have the same variance, i.e., $\sigma_{01s} = 0$ and $\sigma^2_{0s} = \sigma^2_{1qs} = 1$ $\forall s$, and that the random intercept and slopes of 6 of the 30 biomarkers contribute to determining the survival times, i.e., $|\gamma_{s}|, |\delta_{s}| \sim  U(0.5, 1)$ for $s \in \{1, 2, ..., 6\}$, and $\gamma_{s}, \delta_{s} = 0$ otherwise, in equation \eqref{eq:prc-lmm}. Thus, in simulation 1 the random intercepts and slopes play an equal role in determining the time-to-event outcome. In simulation 2 and 3, instead, we consider cases where the slopes play a more prominent role. In particular, in simulation 2 we still let $\sigma^2_{0qs} = \sigma^2_{1qs} = 1$ $\forall q, s$, but we assume that only the random slopes have non-null coefficients in equation \eqref{eq:prc-lmm}. Instead, in simulation 3 we set $\sigma^2_{0qs} = 0.1$ and $\sigma^2_{1qs} = 2$ $\forall q, s$, and similar to in simulation 1 we assume that both the random intercepts and random slopes of 6 biomarkers contribute to determining the survival times. Within each simulation, we consider two different sample sizes ($n = 100$ and $n = 300$). We compute the optimism-corrected tdAUC (with $t = 5$) and C index using bootstrap samples of size $B = 100$, and repeat each simulation setting 100 times. Hereafter we present the results obtained using the ridge penalty, and in the Supplementary Material we show the equivalent results when the elasticnet penalty is used.

Figure \ref{fig:sim1} show the distribution of the tdAUC and C index in simulations 1, 2 and 3 for the baseline pCox model, and for the PRC LMM model when few or many repeated measurements are available. In each of the simulation scenarios we can observe that use of the PRC LMM produces an improvement over the baseline pCox; in particular, the improvement is stronger with the larger sample size ($n = 300$), and when more repeated measurements are available. Furthermore, the improvement is somewhat stronger in simulations 2 and 3, where the survival times are more strongly determined by the progression rates of the biomarkers. 

Similar results are obtained when using elasticnet instead of ridge as penalty (Supplementary Figure 1). We can further observe that the distribution of the performance measures across random replicates of the same simulation has somewhat higher variability with elasticnet than with ridge. This result is probably due to the fact that the elasticnet penalty is a linear combination of an $L^1$ and an $L^2$ penalty that comprises two tuning parameters, $\alpha \in [0, 1]$ and $\lambda > 0$. When $\alpha = 0$ elasticnet reduces to ridge, and when $\alpha = 1$ to the lasso. Because with elasticnet the cross-validation of the tuning parameters can yield different choices of $\alpha$ across random replicates whereas with ridge $\alpha$ is fixed, we can indeed expect the distribution of performance measures to be more variable with elasticnet than with ridge.

Next, we consider the same simulation scenarios in a high-dimensional setting with $p = 150$ predictors. We assume that survival time depends on the first 10 predictors, and in simulations 4, 5 and 6 we make the same assumptions on the variances of the random effects and on the relationship between random effects and survival times made in simulations 1, 2 and 3. The results of these simulations using the ridge penalty are presented in Figure \ref{fig:sim4}. When we consider the optimism-corrected tdAUC estimates, the results are largely in line with those obtained in the low-dimensional simulations: the gain associated to the use of the PRC LMM model is stronger with larger sample sizes, when more repeated measurements are available, and when (in simulations 5 and 6) the survival times are more strongly associated with the random slopes. When looking at the C index, instead, we see that the gain associated with the PRC-LMM is apparent with $n = 300$, but not with $n = 100$. This latter result seems to indicate that obtaining performance improvements can be more challenging in a high-dimensional setting, and that larger sample sizes might be required to accurately evaluate predictiveness in high-dimensions. 

Although similar remarks hold for elasticnet (Supplementary Figure 2), once again we observe a higher variability in the distribution of the performance measurements; furthermore, while the improvement with the PRC LMM method is still apparent in terms of tdAUC, it tends to vanish when looking at the C index.

The primary motivation for the development of PRC is that estimation of joint models becomes computationally unfeasible when more than a handful of longitudinal predictors are available. As a matter of fact, PRC represents a computationally feasible alternative to joint models for the prediction of survival from high-dimensional longitudinal predictors. Nevertheless, it might be still be interesting to compare the predictive performance of PRC to that of joint models \textit{in low-dimensional settings}, where both approaches can be pursued. In Section 3 of the Supplementary material, we present the results of a simulation study designed to compare the performance of PRC LMM to a joint modelling approach in a situation with only 3 longitudinal predictors.

\subsection{Evaluation of the PRC MLPMM(U) and PRC MLPMM(U+B) methods}
\label{sub:mlpmmsims}

\begin{figure}
\centering
\includegraphics[scale=0.35, page = 1]{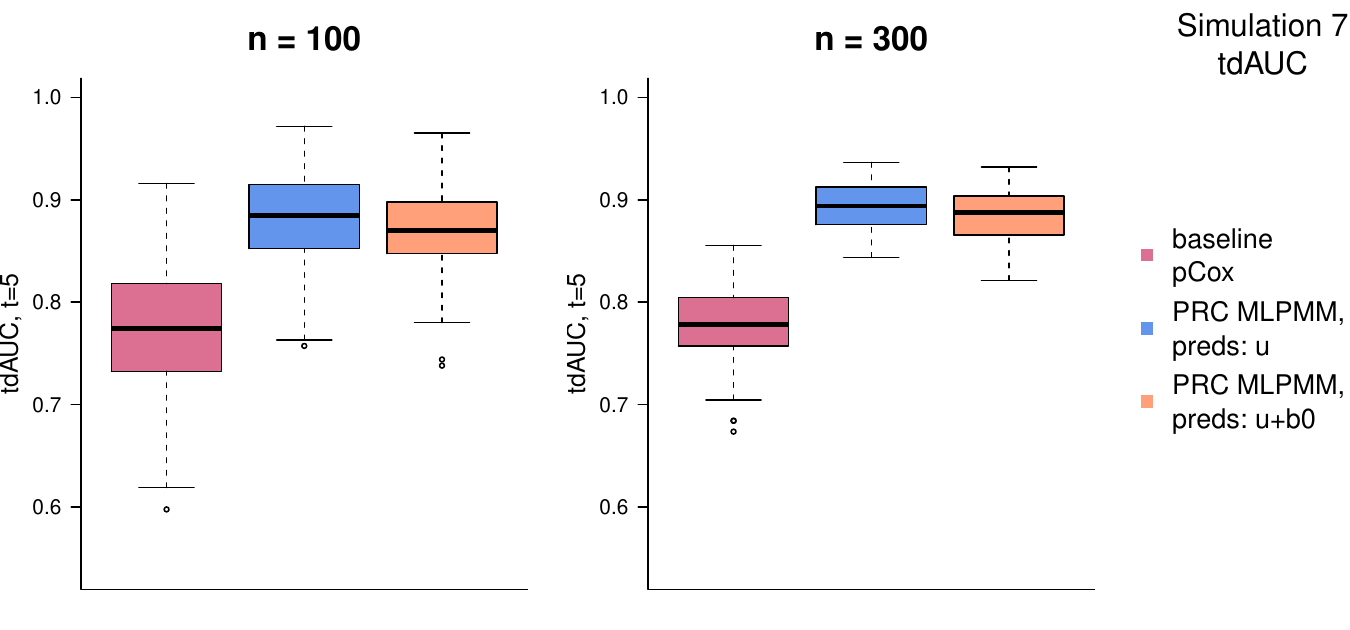}
\includegraphics[scale=0.35, page = 1]{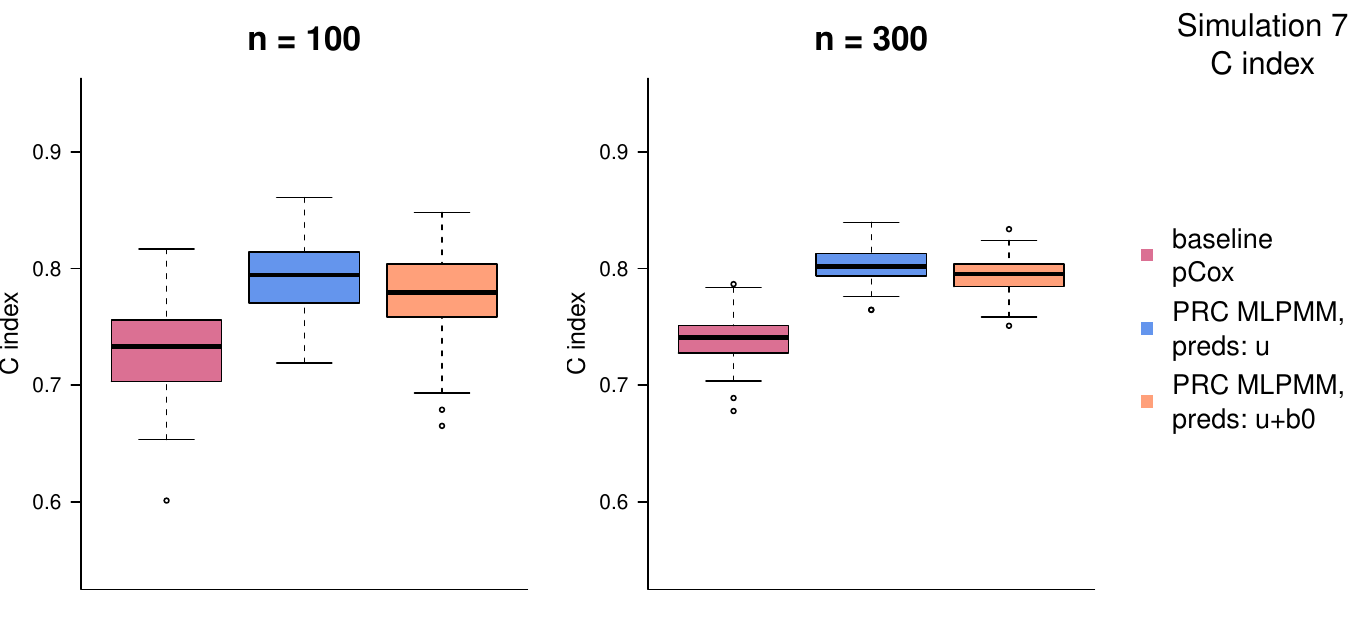}\\
\includegraphics[scale=0.35, page = 2]{tdAUC_multiv_ridge_lowd.pdf}
\includegraphics[scale=0.35, page = 2]{Cindex_multiv_ridge_lowd.pdf}\\
\includegraphics[scale=0.35, page = 3]{tdAUC_multiv_ridge_lowd.pdf}
\includegraphics[scale=0.35, page = 3]{Cindex_multiv_ridge_lowd.pdf}
\caption{Results of simulations 7, 8 and 9 (where we consider a low-dimensional setting with 30 items targeting 10 different biological processes) using the ridge penalty. The boxplots compare the distribution over 100 random replications of the optimism-corrected tdAUC (left) and C index (right) of the baseline pCox (violet red),  PRC-MLPMM(U) (blue) and PRC-MLPMM(U+B) (orange) models.}
\label{fig:sim7}
\end{figure}

\begin{figure}
\centering
\includegraphics[scale=0.35, page = 1]{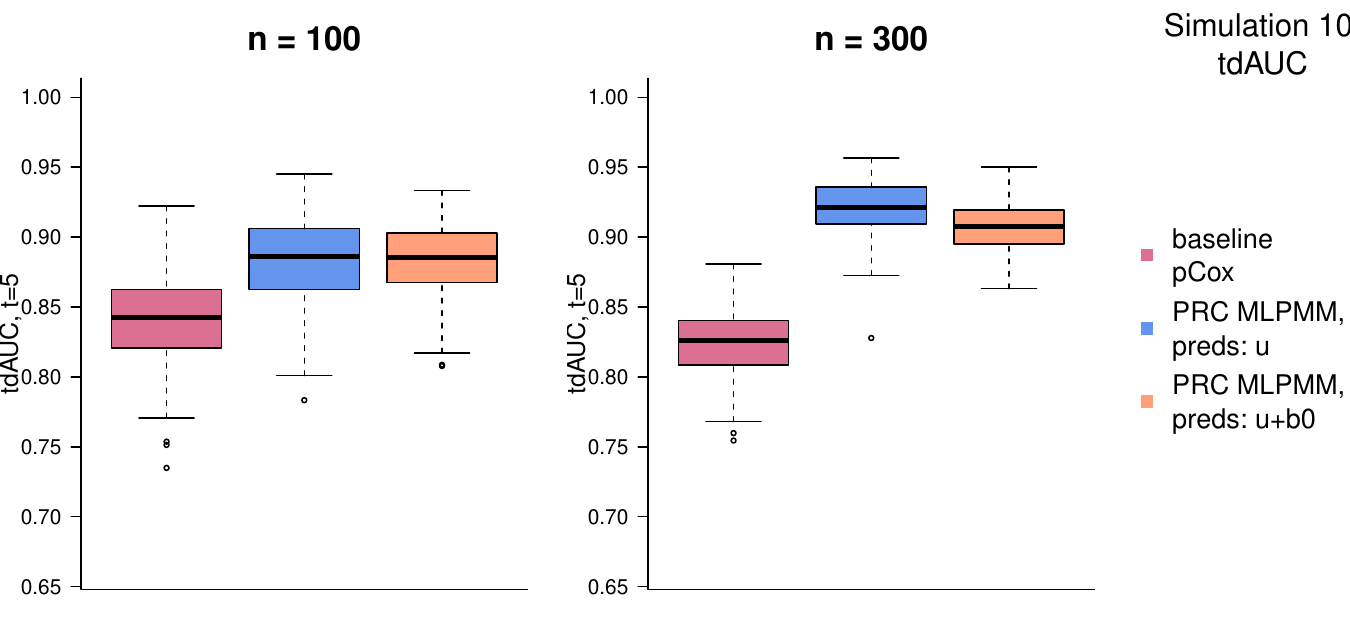}
\includegraphics[scale=0.35, page = 1]{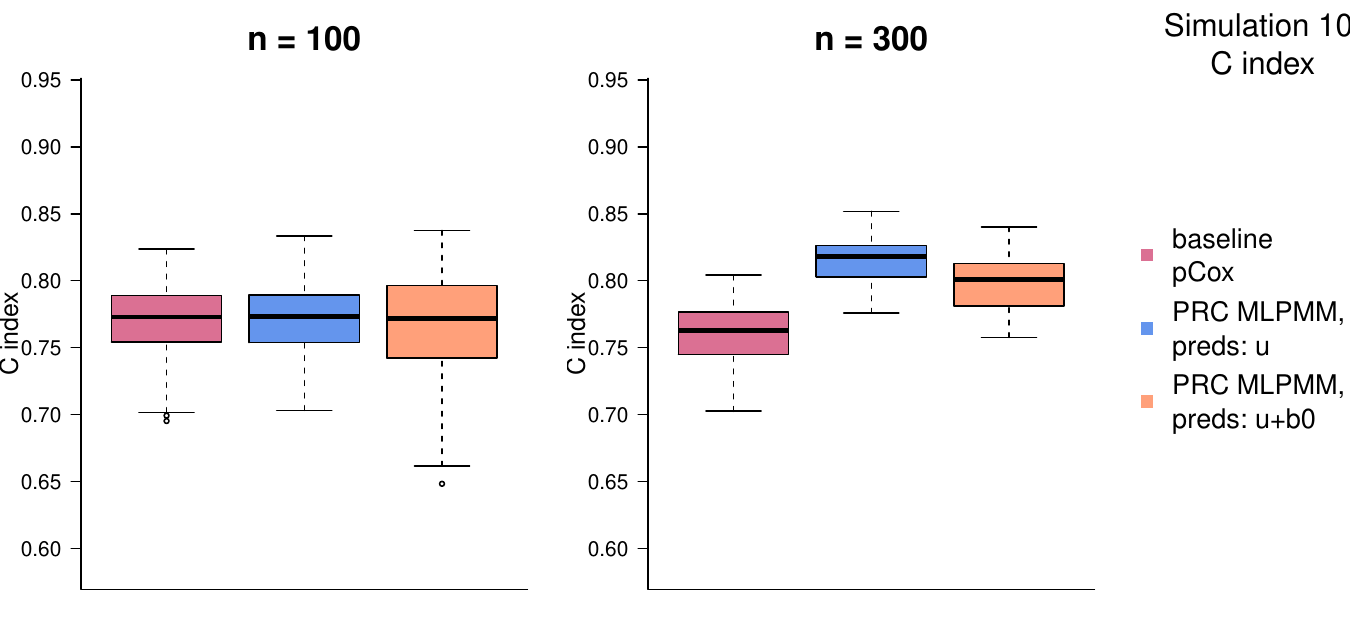}\\
\includegraphics[scale=0.35, page = 2]{tdAUC_multiv_ridge_highd.pdf}
\includegraphics[scale=0.35, page = 2]{Cindex_multiv_ridge_highd.pdf}\\
\includegraphics[scale=0.35, page = 3]{tdAUC_multiv_ridge_highd.pdf}
\includegraphics[scale=0.35, page = 3]{Cindex_multiv_ridge_highd.pdf}
\caption{Results of simulations 10, 11 and 12 (where we consider a low-dimensional setting with 150 items targeting 50 different biological processes) using the ridge penalty. The boxplots compare the distribution over 100 random replications of the optimism-corrected tdAUC (left) and C index (right) of the baseline pCox (violet red),  PRC-MLPMM(U) (blue) and PRC-MLPMM(U+B) (orange) models.}
\label{fig:sim10}
\end{figure}

We now turn our attention to the validation of the PRC MLPMM(U) and PRC MLPMM(U+B) methods, considering simulations where we have multiple correlated items that refer to the same latent biological process. 

Similarly to what we have done in Section \ref{sub:lmmsims}, we begin by considering a low-dimensional scenario where there are $p = 10$ latent biological processes, each measured through $r_s = 3$ items. We generate the longitudinal items using the MLPMM of equation \eqref{eq:mlcmm}, and survival times using a Weibull model where survival depends on the shared random effects, i.e. on $u_{s0}, u_{s1}, s = 1, ..., p$. In simulation 7 we set $\Sigma_{us} = \left[ \begin{array}{cc} 1 & 0.5\\ 0.5 & 1 \end{array} \right] \forall s$ and let survival depend on the shared random effects of 4 of the 10 latent biological processes; in other words, random intercepts and random slopes have the same weight both in the definition of the longitudinal profiles and in the generation of survival times (note that this is a multivariate equivalent of simulations 1 and 4). In simulation 8 we keep the same $\Sigma_{us}$, but we let survival depend only on the random slopes of the latent processes (in analogy with simulations 2 and 5). Lastly, in simulation 9 we set $\Sigma_{us} = \left[ \begin{array}{cc} 0.1 & 0.5\sqrt{0.2}\\ 0.5\sqrt{0.2} & 2 \end{array} \right] \forall s$, and we let survival depend both on the shared random intercepts and on the shared random slopes of the latent processes (this is the equivalent of simulations 3 and 6). Within each simulation, we consider two different sample sizes ($n = 100$ and $n = 300$). We compute the optimism-corrected tdAUC at $t = 5$ and the C index using bootstrap samples of size 100, and for each simulation setting we consider 100 random replicates. Here we present the results obtained using the ridge penalty, and in the Supplementary Material the equivalent results when using elasticnet.

The distributions of the tdAUC and C index in simulations 7, 8 and 9 using the ridge penalty are presented in Figure \ref{fig:sim7}. In all simulations we observe that the two PRC MLPMM methods outperform the baseline pCox both with respect to the tdAUC and to the C index, and with both sample sizes. Moreover, we can observe that the PRC-MLCMM(U) and PRC-MLCMM(U+B) models achieve a similar predictive performance, despite the fact that the data are generated in line with the PRC-MLCMM(U) model, and not with the PRC MLPMM(U+B); as a matter of fact, PRC-MLCMM(U) performs slightly better, as expected given the way in which the survival times are simulated. Supplementary Figure 3 shows the results of simulations 7, 8 and 9 using the elasticnet penalty; in this case, the results are substantially the same as those obtained with ridge.

Lastly, we assess the same simulation scenarios in a high-dimensional setting with $\sum_{s=1}^p \sum_{q=1}^{r_s} = 150$ predictors, where in particular there are $p = 50$ latent biological processes, each measured using 3 items. We assume that survival time depends on the shared random effects of the first 10 latent variables, and in simulations 10, 11 and 12 we make the same assumptions on the variances of the random effects and on the relationship between random effects and survival times made in simulations 7, 8 and 9. 

The results of these simulations using the ridge penalty are presented in Figure \ref{fig:sim10}. Similarly to what we observed in simulations 4, 5 and 6 (in Section \ref{sub:lmmsims}), the improvement of the PRC methods over the baseline pCox is apparent with both $n = 100$ and $n = 300$ when looking at the tdAUC, and with $n = 300$ the C index; it is less apparent when we consider the C index in the $n = 100$ scenarios. Moreover, in line with what already observed in simulations 7, 8 and 9, the difference between PRC-MLPMM(U) and PRC-MLPMM(U+B) is relatively small. Similar remarks hold for elasticnet (Supplementary Figure 4), where, however, we can once again observe a higher variability in the distribution of the performance measurements, similar to simulations 4-6.

\subsection{Effect of the cluster bootstrap optimism correction}
\label{sub:naivevscboc}

To showcase the effect that the CBOCP has on the level of the reported measures of predictiveness, we have compared the distribution of the naive estimates of the tdAUC and C index to that of the optimism-corrected estimates obtained with the CBOCP introduced in Section \ref{sub:bootstrap}. 

\begin{figure}
\centering
\includegraphics[scale=0.55, page = 1]{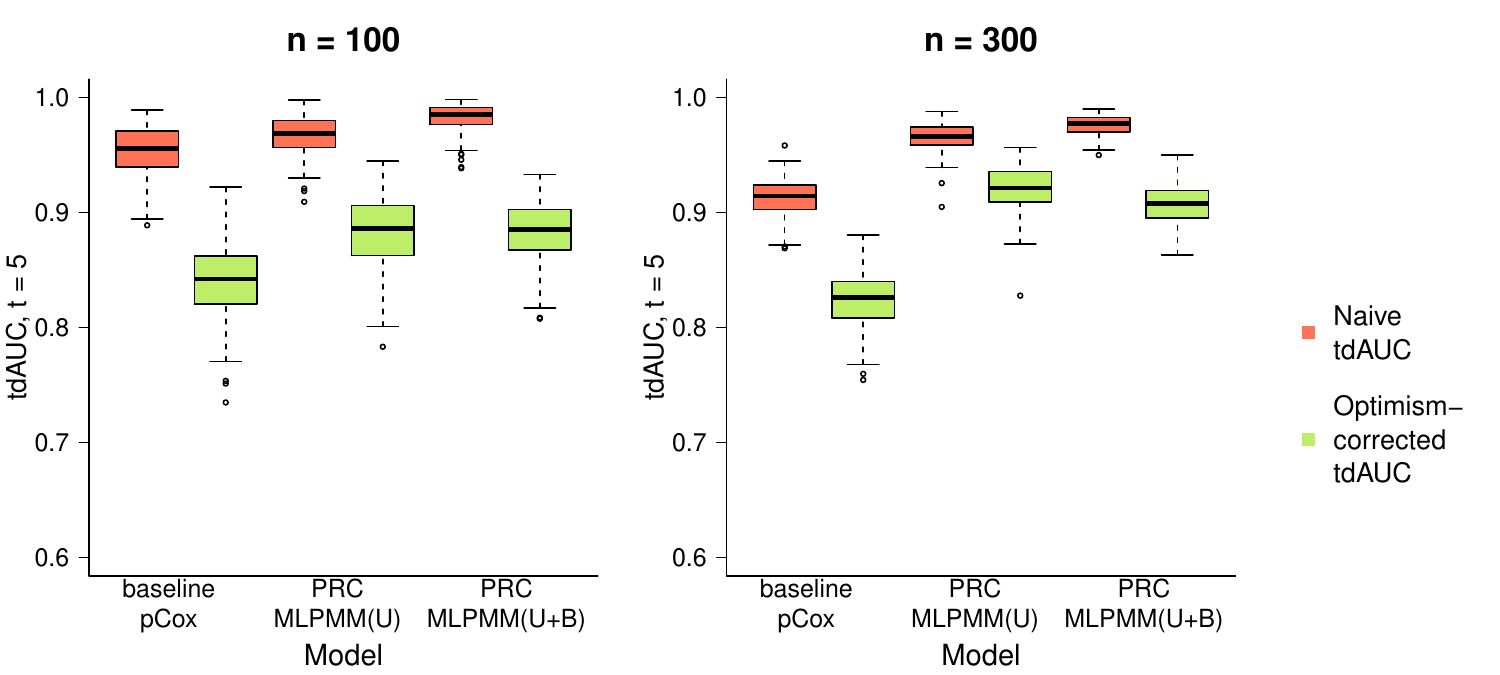}\\
\includegraphics[scale=0.55, page = 2]{0.3-effect_optimism_correction}
\caption{Comparison of the distribution of the naive (orange) and optimism-corrected (green) performance measures in simulation 10. The boxplots compare the distribution over 100 random replications of the naive and optimism-corrected estimates of the tdAUC (top panel) and C index (bottom) for each of the methods considered in simulation 10.}
\label{fig:cbocp}
\end{figure}

Figure \ref{fig:cbocp} compares the naive and optimism-corrected estimates of the tdAUC and C index in simulation 10 (similar results hold for the other simulations described in Sections \ref{sub:lmmsims} and \ref{sub:mlpmmsims}). It is apparent that the effect of the optimism correction is rather sizeable for all models. For example, in the case of the PRC-MLPMM(U+B) model with $n = 300$ the median of the naive estimates is 0.978 for tdAUC(5) and 0.881 for the C index, whereas the median of the optimism-corrected estimates is 0.908 for tdAUC(5) and 0.801 for the C index. 
These results highlight the importance of properly evaluating predictive performance by implementing an internal validation procedure that can estimate the optimism associated to the naive performance estimators.

\subsection{Computing time}
\label{sub:comptime}

In this Section we study the typical computing time of the PRC-LMM and PRC-MLPMM models and its relation to sample size and number of longitudinal predictors. For this benchmarking, we employ the data generated in simulations 7 and 10, and monitor the mean computing time of a penalized Cox model with baseline predictors, of the PRC-LMM model, and of the PRC-MLPMM(U) and PRC-MLPMM(U+B)) models. Computations were performed using an Intel E7-4890 processor with 2.2 GhZ CPU, and version 0.4.1 of the \texttt{R} package \texttt{pencal}; for each model, computation of the CBOCP was parallelized over 10 cores. 

Supplementary Table 1 reports the computing time associated to the estimation of the different models. We can observe that both for the PRC-LMM and the PRC-MLPMM models, computing time increases less than linearly with respect to the sample size $n$. Moreover, computing time increases more than linearly with respect to the number of longitudinal predictors for the PRC-LMM approach, and approximately linearly for the two PRC-MLPMM approaches. Similar observations hold for the computation of the CBOCP, whose mean computing times are presented in Supplementary Table 2.

By comparing the computing times in Supplementary Tables 1 and 2, we can observe that while the estimation of PRC is not particular demanding in terms of computing time, the computation of the CBOCP is typically slower. This is due to the fact that bootstrap validation procedures require the repetition of the same computations for a rather large number (typically hundreds) of bootstrap samples. 
Due to their repetitiveness, these computations can be easily parallelized: for this reason, in the \texttt{R} package \texttt{pencal} we offer the possibility to easily parallelize computations over multiple cores, making it easy for users to reduce computing time through parallelization.

\section{Application}
\label{sec:appl}

In this section we illustrate an example application of PRC to the data gathered in the MARK-MD study, the motivating dataset that we briefly introduced in Section \ref{sec:intro}. The MARK-MD study \citep{signorelli2020,strandberg2020} was an observational study that focused on the validation of a set of blood biomarkers as longitudinal markers of disease progression for Duchenne muscular dystrophy (DMD), a severe neuromuscular disorder whose consequences include progressive loss of muscular tissue and muscle mass, loss of ambulation (LoA) around the age of 12, and premature death.
One of the reasons that motivate the research interest in blood biomarkers for DMD is that the methods currently used to monitor disease progression in DMD (namely, timed tests that measure the distance and speed that a patient can walk, and magnetic resonance imaging) can be motivation-dependent and rather noisy, so that measuring the level and dynamic change of blood biomarkers would represent a desirable alternative.

\begin{figure}
\centering
\includegraphics[scale=0.5]{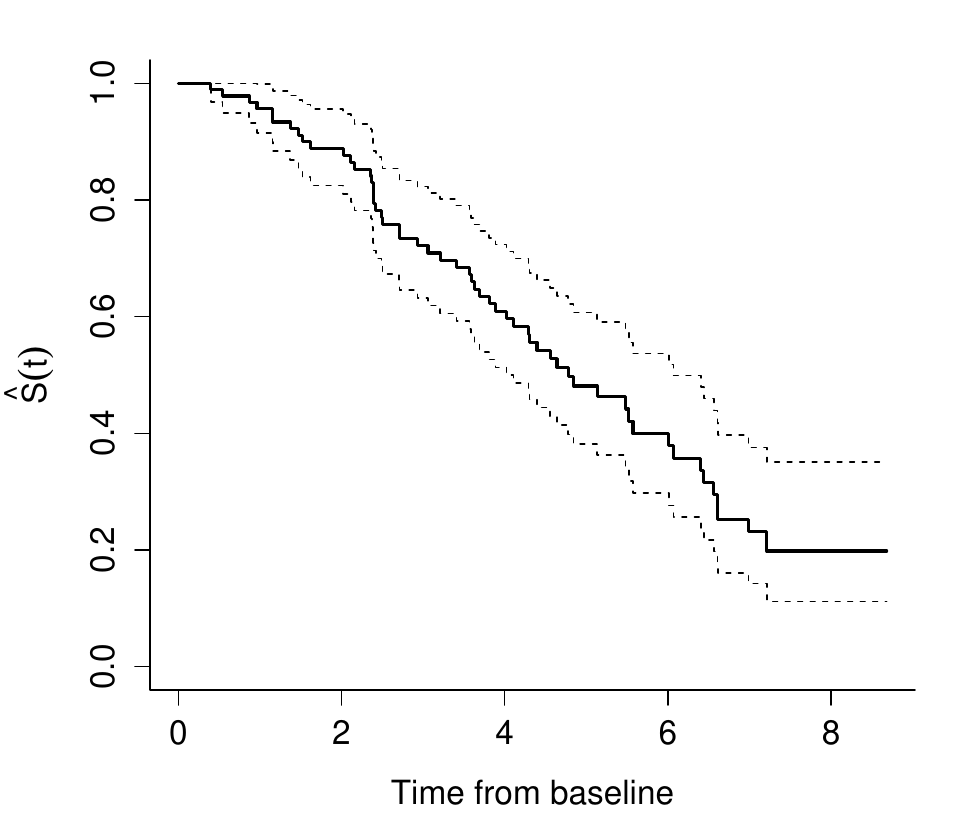}
\caption{Kaplan-Meier estimator for the MARK-MD data. The estimated survival probabilities $\hat{S}(t)$ correspond to the probability of being ambulant after $t$ years from baseline.}
\label{fig:kaplan}
\end{figure}

Hereafter we focus on the problem of predicting the age at which DMD patients will experience LoA, which is a crucial disease milestone for DMD. The MARK-MD study led to the collection of 303 longitudinal blood samples from a total of 157 patients. However, 64 of these patients were not ambulant where they entered the study, so here we restrict our attention to the 93 patients who were ambulant when they entered the MARK-MD study. Of these 93 patients, 55 experienced the event of interest (LoA), whereas 38 are right-censored. Figure \ref{fig:kaplan} shows the Kaplan-Meier estimator associated to the LoA event.
\begin{figure}
\centering
\includegraphics[scale=0.55	]{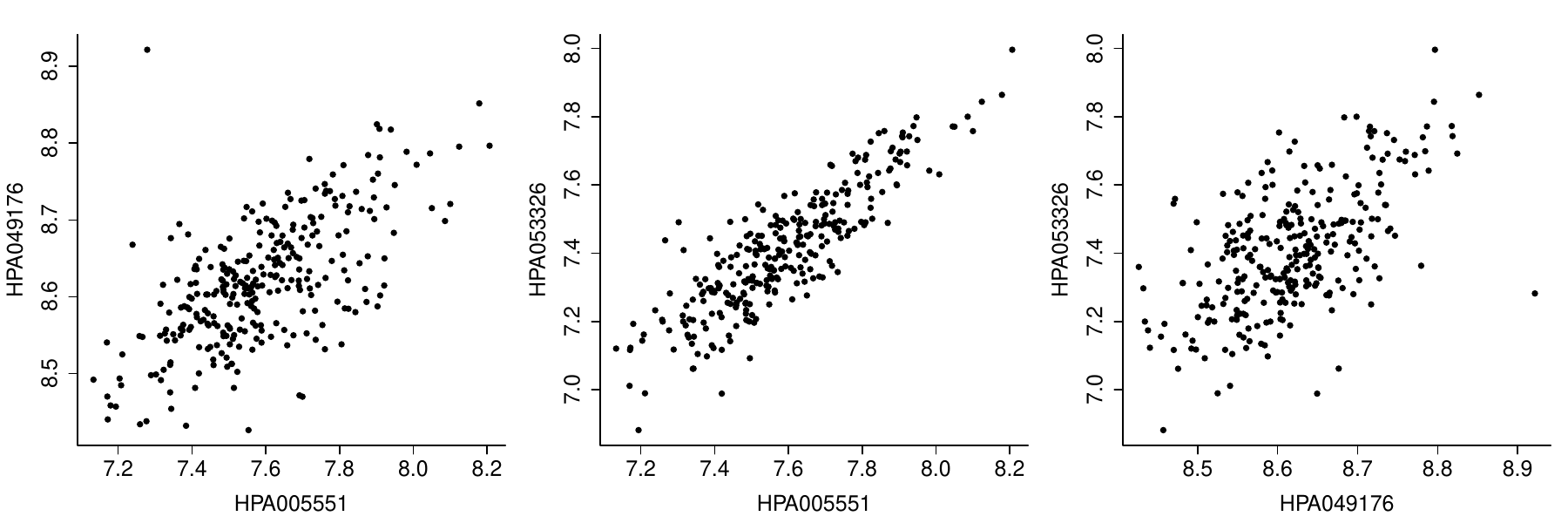}
\caption{Correlation between the three items employed to measure the protein CFH. In the MARK-MD study, the protein CFH was measured using three antibodies: HPA005551, HPA049176 and HPA053326. The three scatter plots compare each pair of antibodies, showing that they are all positively correlated with each other.}
\label{fig:cfh}
\end{figure}
\begin{center}
\begin{table*}
\caption{Distribution of the number of repeated measurements before LoA across patients.\label{tab:num-rep-meas}}
\vspace{0.5cm}
\centering
\begin{tabular}{l|ccccc}
\hline
Number of repeated measurements before LoA & 1 & 2 & 3 & 4 & 5\\
\hline
Frequency & 45 & 23 & 18 & 6 &  1\\
\hline
\end{tabular}
\end{table*}
\end{center}
An interesting feature of the MARK-MD study is the fact that the biological markers of interest, i.e. the proteins, were not measured directly; instead, between 1 and 5 antibodies, with different affinity and specificity towards a given target protein, were employed to measure each protein.
More in detail, an antibody-based array was employed to measure the abundance of 240 antibodies that targeted 118 proteins in total; 37 proteins were measured using a single antibody, 52 using two different antibodies, 18 with 3 distinct antibodies, 10 with 4 antibodies, and 1 with 5 different antibodies. The 118 proteins partitioned the antibodies into non-overlapping sets (i.e., each antibody targeted a single protein).
In other words, we are interested in $p = 118$ latent biological process, each of which is measured using between 1 and 5 items, i.e. $r_s \in \{1, 2, 3, 4, 5\}$. 

In most cases, a moderate / strong positive correlation can be observed between antibodies that target the same protein (see Figure \ref{fig:cfh} for an example based on the three antibodies used to measure the protein CFH, and Supplementary File 1 of the original study \citep{signorelli2020} for a more detailed overview), indicating that a multivariate modelling of the antibodies that match the same protein might be more appropriate than modelling each antibody separately.

Table \ref{tab:num-rep-meas} shows the number of repeated blood measurements taken before LoA that are available for each patient. Note that while between 2 and 5 repeated measurements before LoA are available for 48 patients, for 45 patients only 1 measurement can be used: this feature makes the longitudinal modelling of the biomarkers challenging, because the fact that no longitudinal information is available for almost half of the patients makes it is difficult to derive accurate predicted random effects, and in particular to obtain accurate random slopes.

We estimated four different models that can be employed to predict time to LoA: a penalized Cox model that employs the baseline biomarkers levels as covariates, the PRC-LMM model that in step 1 models each antibody separately with a LMM with random intercepts and random slopes, and the PRC-MLCMM(U) and PRC-MLCMM(U+B) models, which jointly model all antibodies targeting the same protein using the MLPMM. Each of the models was fitted using both the ridge, and the elasticnet penalty. The predictive accuracy of each model was evaluated by estimating the optimism-corrected C index, and the optimism-corrected tdAUC evaluated every half year, up to 5 years from baseline. The optimism correction was performed using the CBOCP with $B = 500$ bootstrap samples.

The optimism-corrected C index estimates of the aforementioned models are compared in Table \ref{tab:cindex}. Irrespective of the type of penalty used, the best performing model appears to be the PRC-MLPMM(U+B). It is interesting to note that whereas this model outperforms the penalized Cox model that uses only the baseline marker values, the PRC-LMM and PRC-MLPMM(U) models don't. 
We believe that this somewhat surprising result may be due to the lack of repeated measurements for almost half of the patients: as a matter of fact, for those 45 patients for which only 1 measurement before LoA is available it is virtually impossible to reliably estimate the random slopes (progression rates). 
We can expect this complication to affect more severely the performance of the PRC-LMM and PRC-MLPMM(U) models than that of the PRC-MLPMM(U+B) model, because whereas in the PRC-LMM and PRC-MLPMM(U) models half of the predictors are random intercepts and half random slopes, in the PRC-MLPMM(U+B) the random intercepts ($b_{qsi}$ and $u_{s0i}$ in equation \eqref{eq:mlcmm}) employed for prediction outnumber the random slopes ($u_{s1i}$; the total number of random intercepts is 237, versus 118 random slopes).

\begin{figure}
\centering
\includegraphics[scale=0.55]{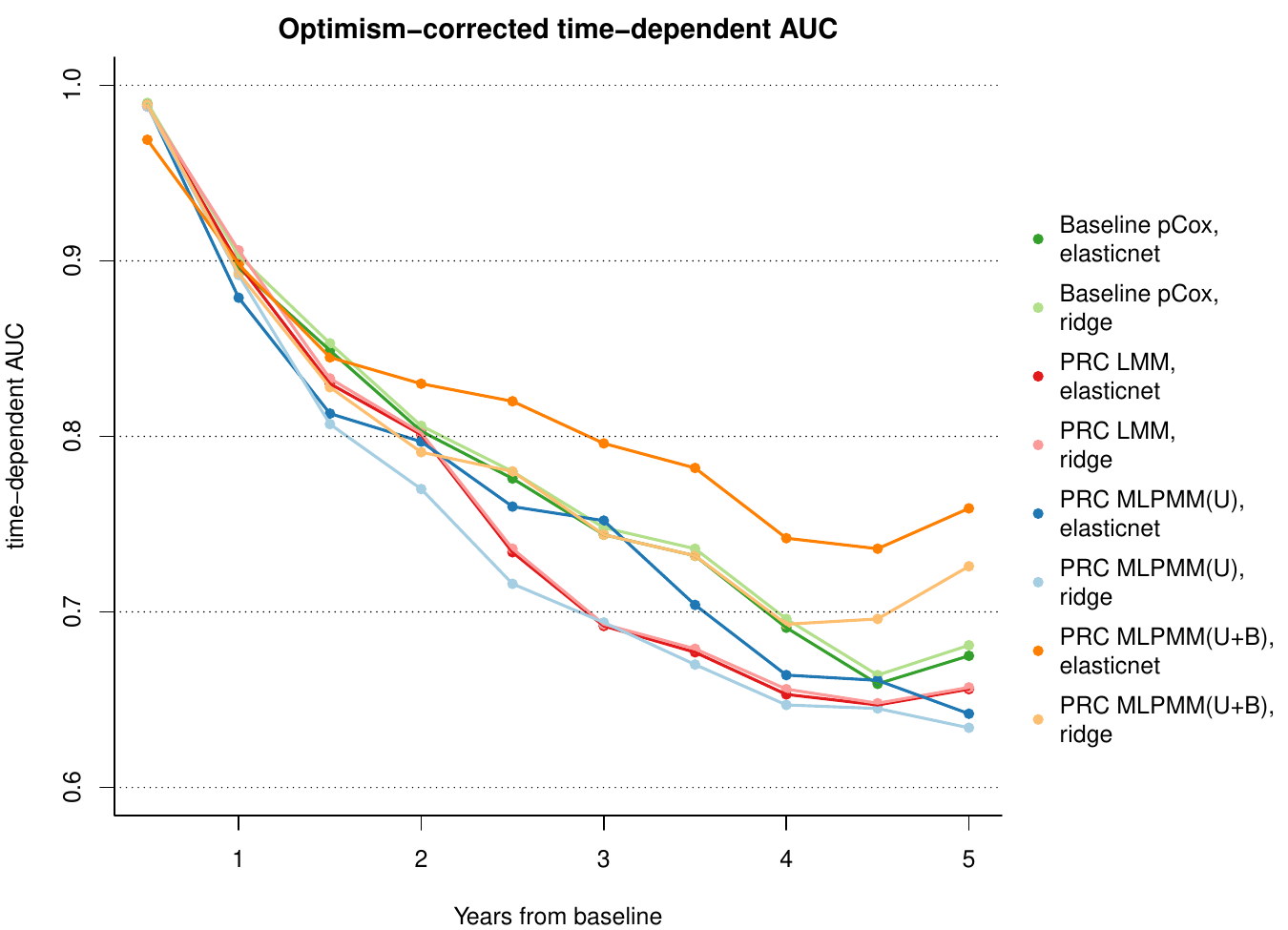}
\caption{Optimism-corrected estimates of the tdAUC for the prediction of time to LoA in the MARK-MD dataset.}
\label{fig:tdauc}
\end{figure}

Similar results hold for the optimism-corrected estimates of the tdAUC, which are shown in Figure \ref{fig:tdauc}. Once again, the most predictive model appears to be the PRC-MLPMM(U+B) model when fitted with elasticnet penalty, followed by the same model fitted with the ridge penalty and  by the penalized Cox model with baseline measures as covariates.

Lastly, we note that for the baseline penalized Cox and the PRC-LMM models, the predictive performance is substantially the same when using ridge or elastic net; this is due to the fact that for these two models, the value of the elastic net tuning parameter $\alpha$ on the original dataset selected by nested cross-validation is $\hat{\alpha}_{NCV} = 0$, meaning that the elastic net model is the same as the ridge model. Thus, the naive estimates of the C index and tdAUC are the same for the two different penalties. Slight differences in the estimation of the optimism correction are however expected, because with elasticnet, $\alpha$ is tuned independently for each boostrap sample during the CBOCP. This, in turn, yields slightly different optimism-corrected estimates of performance. On the contrary, for the PRC-MLPMM(U) and PRC-MLPMM(U+B) models we observe a more substantial difference between ridge and elastic net: this is due to the fact that for these models, $\hat{\alpha}_{NCV} > 0$, so the models selected by ridge and elastic net are actually different, and have different predictive performance.
\begin{center}
\begin{table*}[t]%
\caption{Optimism-corrected estimates of the C index for the prediction of time to LoA in the MARK-MD dataset.\label{tab:cindex}}
\vspace{0.5cm}
\centering
\begin{tabular}{l|ccccc}
\hline
& \multicolumn{2}{c}{Optimism-corrected C index}\\
Model & Ridge penalty & Elasticnet penalty\\
\hline
Penalized Cox with baseline markers & 0.696 & 0.692\\
PRC-LMM & 0.668 & 0.666\\
PRC-MLPMM(U) & 0.640 & 0.659\\
PRC-MLPMM(U+B) & 0.710 & 0.733\\
\hline
\end{tabular}
\end{table*}
\end{center}

\section{Discussion}
\label{sec:concl}

High-dimensionality and repeated measurements are commonly encountered features of modern biomedical data. However, the current lack of statistical methods that can predict survival using a high-dimensional set of longitudinally measured predictors limits the possibility to exploit the full predictive potential of complex biomedical datasets. 
When dealing with a large number of longitudinal predictors, this methodological gap makes it necessary to employ simpler prediction strategies that are either based on the pre-selection of a limited number of longitudinal covariates as predictors, or on the use of the baseline measurements only. Such simplifications can be detrimental for the performance of risk prediction models, because they discard part of the available information. 

In this paper we overcome this limitation by proposing PRC, a method that makes it possible to predict survival using covariates that are at the same time longitudinal and high-dimensional. 
PRC begins with the specification of mixed effects models that describe the dynamic evolution of the biomarkers, it proceeds to the computation of subject-specific summaries of the biomarker's trajectories, and finally it employs such summaries as predictors of the survival outcome (Figure \ref{fig:prc_diagram}).
Use of mixed effects models makes it easy to handle unbalanced repeated measurement designs, as well as missing data.
Additionally, PRC is capable of handling the presence of multiple items that are employed to measure the same underlying biological process, and which might be highly correlated with each other.

To illustrate PRC we have employed data from the MARK-MD study as motivating example. In this study, the measured predictors are antibodies that are employed to reconstruct the abundance of a smaller number of proteins, which cannot be measured directly and can thus be regarded as the latent biological processes of interest. 
To account for this peculiar feature, we have introduced PRC considering a general framework where multiple items can be employed to measure the same biological process of interest.
However, here we would like to emphasize that the applicability of PRC is not limited to studies with such a complex setup: PRC can also be applied to simpler contexts where no distinction exists between the items and the biological processes of interest. In such situations, use of the PRC LMM approach instead of the PRC MLPMM might be preferable.

Moreover, we would like to remark that although we have illustrated PRC using an example where proteomic data are employed as predictors, the applicability of PRC is not restricted to proteomic datasets. As a matter of fact, there is no restriction on the type of longitudinal variables that might be used, so that in practice PRC can be applied to datasets comprising a highly heterogeneous set of variables, which might for example include clinical variables, laboratory measurements, different types of omic measurements, and performance tests.
Moreover, applicability of PRC is not restricted to problems arising in medicine and biology, as the method can be employed with any type of survival outcomes and longitudinal predictors.

To estimate the penalized Cox models in step 3 of PRC (Section \ref{sub:prc-step3}) we have focused on the use of the ridge and elasticnet penalties, which are commonly employed for prediction problems. If desired, one may consider the use of alternative forms of regularization (lasso, group lasso, SCAD, ...). In the case of equation \eqref{eq:elnet}, one might also consider a modification of elasticnet that combines the ridge and group lasso penalty to jointly select or exclude all summaries that refer to a single biological process.
Although in the \texttt{R} package \texttt{pencal} we allow for the possibility to use the lasso penalty, in general we do not recommend its use, because lasso models typically have lower predictive performance in comparison to ridge and elastic net, and they additionally have problems when dealing with highly-correlated predictors \citep{zou2005}.

In the simulations presented in Section \ref{sec:sims} we observed that ridge and elasticnet usually result in a similar prediction performance; however, the distribution of the optimism-corrected tdAUC and C index over randomly replicated datasets is often more variable with elasticnet, indicating that use of ridge may yield more stable predictors. 
Moreover, the estimation of PRC is faster when employing the ridge penalty, because this is a special case of elasticnet. Thus, our practical recommendation is to estimate PRC using the ridge penalty, unless a sparse predictor is desired.

A further contribution of our work is the implementation of the CBOCP, which makes it possible to perform a proper internal validation of the fitted prediction model. As shown in Section \ref{sub:naivevscboc}, the optimism correction in high-dimensional settings can be substantial, indicating that the naive measures of predictive performance may exhibit a strong bias, and that performing the CBOCP is thus necessary to correct for this bias.

Motivated by the presence in the MARK-MD study of strong correlations between antibodies that target the same protein, in this article we have proposed a multivariate modelling approach where all items targeting the same biological process are modelled jointly (step 1 of the PRC-MLPMM model).
Primary advantages of this approach are the facts that it can tackle the major source of correlation between the longitudinal antibodies, and that it allows to obtain summaries of the longitudinal trajectories that refer to the latent biological processes. On the other hand, this approach does not model the weaker correlations that exist between antibodies that target different proteins, as specifying a single multivariate mixed model for 240 longitudinal biomarkers is computationally prohibitive. If computationally feasible, an extension of PRC-MLPMM whereby all 240 longitudinal biomarkers are jointly modelled in step 1 could yield prediction models with higher predictive accuracy.

Lastly, in this article we have assumed that survival times have been systematically collected and reported as an outcome measured in continuous time. However, in clinical practice it may happen that the survival status of a patient is only measured at discrete time points, for example on the dates of a visit, resulting in interval censoring. Dealing with interval-censored outcomes would require an extension of PRC whereby the penalized Cox model employed in step 3 of PRC is replaced with a model for interval-censored data suitable for the analysis of high-dimensional data.

\section*{Author contributions}

Mirko Signorelli and Roula Tsonaka conceived this study, developed the methodology described in Section \ref{sec:methods}, planned the simulations presented in Section \ref{sec:sims} and analysed their results. 
Mirko Signorelli developed the \verb|R| package \verb|pencal|, implemented the simulations, and performed the statistical analyses presented in Section \ref{sec:appl}.
Pietro Spitali, Cristina Al-Khalili Szigyarto and the MARK-MD Consortium contributed the biomedical data from the MARK-MD study that are analysed in Section \ref{sec:appl}.
Mirko Signorelli, Pietro Spitali and Roula Tsonaka worked on the application of PRC to the MARK-MD data.
Mirko Signorelli  and Roula Tsonaka wrote the manuscript; all co-authors revised the manuscript.

\section*{Collaborators}

The authors gratefully acknowledge the following collaborators from the MARK-MD Consortium: 
prof. Erik Niks, Department of Neurology, Leiden University Medical Center (NL);
prof. Volker Strauss, MRC Centre for Neuromuscular Diseases, Institute of Genetic Medicine, Newcastle University (UK);
prof. Francesco Muntoni, The Dubowitz Neuromuscular Centre, UCL Institute of Child Health (UK);
dr. Burcu Ayoglu, Department of Protein Science, KTH-Royal Institute of Technology (SW);
dr. Camilla Johansson, Department of Protein Science, KTH-Royal Institute of Technology (SW).

\section*{Acknowledgements}

The authors gratefully acknowledge funding from the Association Francaise Contre les Myopathie (grant number 17724), the Stichting Duchenne Parent Project (project 16.006) and Spieren voor Spieren (grant number SvS15). Moreover, they acknowledge the MRC Centre for Neuromuscular Diseases Biobank London and EuroBioBank for giving access to the samples.

\bibliographystyle{apa}
\bibliography{bibliography}

\clearpage
\newpage


\section*{Supplementary material}

# 3 Comparison to the predictive performance of joint modelling in a low-dimensional settings

The primary motivation for the development of PRC is the fact that the estimation of joint models is computationally intensive, and it is practically unfeasible with more than a handful of longitudinal covariates. In situations where a larger number of longitudinal covariates is available, PRC represents a computationally feasible alternative to joint models for the prediction of survival probabilities. Nevertheless it may be still be interesting to compare the predictive performance of PRC to that of joint models *in low-dimensional settings*, where both approaches can be pursued.

In this Section we compare the predictive performance of PRC to that of joineRML (Hickey et al., 2018), a multivariate joint modelling approach, in a scenario with 3 longitudinal covariates. PRC and joineRML model the relationship between the longitudinal outcomes and the survival time differently, but they are equivalent when the random effects structure comprises only random intercepts. Additionally, joineRML does not offer a built-in strategy to compute optimism-corrected estimates of predictive performance. Therefore, to ensure that the two models are comparable we simulate the longitudinal predictors from LMM models where $y_{sij} = \beta_{s0} + b_{s0i} + (\beta_{s1})a_{ij} + \varepsilon_{sij}$, letting the survival time depend on the random intercepts through a Weibull model. Moreover, we use a split-sample validation approach to estimate predictive performance, with a training set comprising $n = 300$ subjects and a validation set with 200 subjects. joineRML computes conditional survival probabilities from the last available longitudinal measurement; therefore, for simplicity we assume a balanced longitudinal design with 5 longitudinal measurements taken at $t_{ij} = 0, 0.1, 0.3, 0.5, 1$, and that events and censoring occur starting from $t > 1$. We estimate both models on the training set, and we compute the conditional survival probabilities $\hat{S}(t|1)$ for subjects in the validation set.

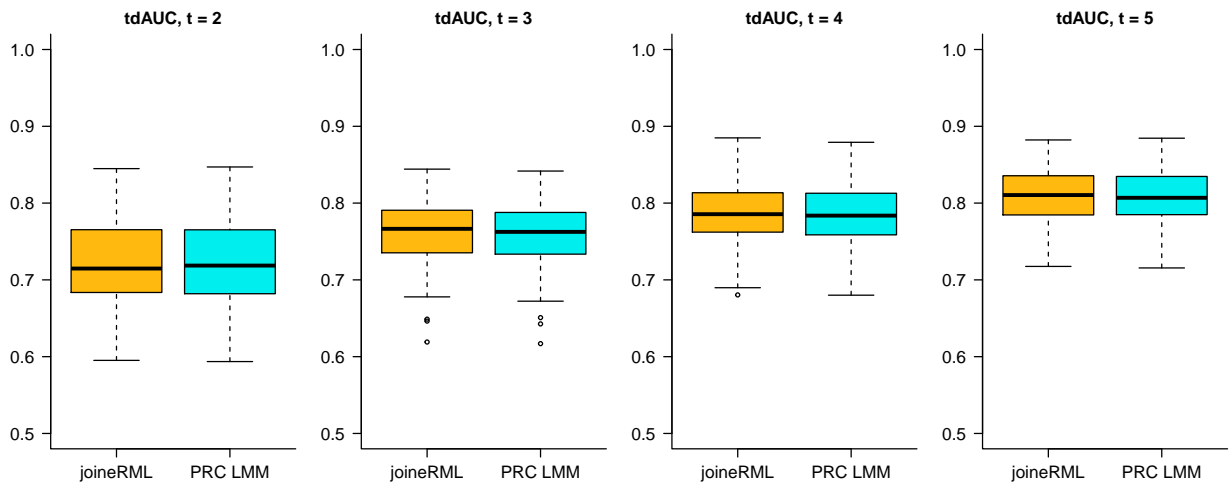

Figure 5: Optimism-corrected time-dependent AUC of joineRML and PRC LMM for predictions at $t = 2, 3, 4, 5$. Results based on 100 replications.

Figure 5 compares the distribution of the optimism-corrected tdAUC for predictions at $t = 2, 3, 4, 5$ for the two models. In short, we can observe that the predictive performance of PRC and joineRML is very similar at each of the considered prediction horizons. It should be noted, however, that estimation of joineRML is considerably slower than that of the PRC LMM: on average, the esti-

mation of PRC LMM took 0.47 seconds, whereas that of joineRML 453.84 seconds (computations were run using a single core on an Intel E7-4890 processor with 2.2 GhZ CPU).



\end{document}